\documentclass[aps,prx,twocolumn,footinbib]{revtex4-2}
\usepackage{graphicx}
\usepackage{indentfirst}
\usepackage{physics}
\usepackage{braket}
\usepackage{float}
\usepackage{amsmath}
\usepackage{epstopdf}
\usepackage{footnote} 
\usepackage{CJK}
\usepackage{esint}
\usepackage{color}
\usepackage[T1]{fontenc}
\usepackage{subfigure}
\usepackage{amsfonts}
\usepackage{footmisc}
\usepackage{scrextend}
\usepackage{multirow}
\usepackage[hyperfootnotes=false]{hyperref}
\usepackage[acronym]{glossaries}

\usepackage[english]{babel}
\usepackage{url}
\usepackage{bm}
\usepackage{hyperref}
\definecolor{darkblue}{rgb}{0,0,0.5}
\hypersetup{
colorlinks=true,
linkcolor=black,
filecolor=blue,
citecolor=darkblue,  
urlcolor=black,
}

\newtheorem{theorem}{Theorem}

\newtheorem{lemma}[theorem]{Lemma}

\newenvironment{proof}[1][Proof]{\noindent\textbf{#1.} }{\ \rule{0.5em}{0.5em}}

\newcommand{\calA}{{\cal A}}
\newcommand{\calC}{{\cal C}}

\newcommand{\calI}{{\cal I}}
\newcommand{\calL}{{\cal L}}

\newcommand{\calH}{{\cal H}}

\newcommand{\calR}{{\cal R}}

\newcommand{\1}{^{(1)}}

\newcommand{\state}[1]{\ketbra{#1}{#1}}

\def\be{\begin{equation}}
\def\ee{\end{equation}}
\def\ba{\begin{eqnarray}}
\def\ea{\end{eqnarray}}
\usepackage{bm}

\newcommand{\QZ}[1]{{{\textcolor{black}{#1}}}}

\begin{document}

\title{\QZ{Quantum-enabled communication without a phase reference}}

\author{Quntao Zhuang}
\email{zhuangquntao@email.arizona.edu}

\address{
Department of Electrical and Computer Engineering \& 
James C. Wyant College of Optical Sciences,
\\
 University of Arizona, Tucson, AZ 85721, USA
}

\begin{abstract}
A phase reference has been a standard requirement in continuous-variable \QZ{quantum} sensing and communication protocols. However, maintaining a phase reference is challenging due to environmental fluctuations, preventing \QZ{quantum phenomena} such as entanglement \QZ{and coherence} from being utilized in many scenarios. We show that \QZ{quantum communication} and entanglement-assisted communication without a phase reference are possible, when a short-time memory effect is present. \QZ{The degradation in the communication rate of classical or quantum information transmission decreases inversely with the correlation time.} An exact solution of the \QZ{quantum capacity} and entanglement-assisted classical/quantum capacity for pure dephasing channels is derived, where non-Gaussian multipartite-entangled states show strict advantages over usual Gaussian sources. For thermal-loss dephasing channels, lower bounds of the capacities are derived. \QZ{The lower bounds also extend to scenarios with fading effect in the channel.} \QZ{In addition, for entanglement-assisted communication, the lower bounds can be achieved by a simple phase-encoding scheme on two-mode squeezed vacuum sources, when the noise is large.}
\end{abstract} 
\maketitle





\QZ{
Quantum physics has re-shaped our understanding of communication. The Shannon capacity has been generalized to the Holevo-Schumacher-Westmoreland (HSW) classical capacity~\cite{hausladen1996classical,schumacher1997sending,holevo1998capacity} to incorporate quantum effects during transmission. 
Entanglement has also enabled non-classical phenomena in communication, such as superadditivity~\cite{hastings2009superadditivity,smith2008quantum,zhu2017,zhu2018superadditivity,leditzky2018,fanizza2020quantum} and capacity-boost from entanglement-assistance (EA)~\cite{bennett1992,bennett1999entanglement,bennett2002entanglement,holevo02,shor2004classical,hsieh2008entanglement,zhuang2017additive,wilde2012quantum,wilde2012information}. Moreover, reliable transmission of quantum information is possible, established by the Lloyd-Shor-Devetak quanutm capacity theorem~\cite{quantum_capacity_Lloyd,quantum_capacity_Shor,quantum_capacity_Devetak}.
}

\QZ{
Apart from the information theoretical advances, the physical realization of quantum-enabled communication protocols, inevitably in the optical domain, have been extensively studied. 
Although the classical capacity has been found additive~\cite{giovannetti2014ultimate}, the quantum capacity of noisy optical communication is still an open question~\cite{rosati2018narrow,sharma2018bounding,noh2018quantum,noh2020enhanced}. In EA communication, the advantage of entanglement has been known for decades, and surprisingly thrives even more in presence of loss and noise~\cite{bennett2002entanglement}. More recently, practical EA classical communication protocols have been designed~\cite{shi2020practical}, and experimentally demonstrated to beat the HSW capacity~\cite{hao2020}.
}


Despite the recent progresses, the need of phase-stabilization---the maintenance of a phase reference between the sender and a receiver---in the above protocols places a serious constraint on their applicability. \QZ{Even in a well-controlled experimental condition~\cite{hao2020}, the instability of phase-locking limits the time-duration of EA communication.} In presence of environmental fluctuations, phase-stabilization over real communication links requires a highly non-trivial feedback control system~\cite{Grein:17}, which might be impossible for wireless scenarios. 



In this letter, surprisingly we show that \QZ{quantum-enabled} communication without a phase reference is possible and advantageous. We adopt the non-Gaussian memory channel model in Ref.~\cite{fanizza2020classical}, where phase fluctuations have a finite timescale such that at most $m$ signal modes can be sent out with a fixed unknown random phase. Despite facing challenges brought by the non-Gaussian nature of the channel, we exactly solve the \QZ{quantum capacity and EA classical/quantum capacity of} such a bosonic dephasing channel. The optimal encoding requires a non-Gaussian multipartite-entangled state and is strictly better than the \QZ{conventional Gaussian encoding} known to be optimal in the absence of phase noise. 


With thermal-loss effects into play, we provide lower and upper bounds of the communication rates, showing that phase noise only decreases the capacity by at most a constant $\sim\log_2(m)/m$ bits per mode. In addition, when the thermal noise in the communication link is high, we show that phase encoding on TMSV achieves the EA capacity lower bounds, assuming \QZ{optimum} receivers for decoding. \QZ{Extending the bounds to channel fading scenarios~\cite{goodman1976some,goodman1965some,zhuang2017fading}, we show that the EA advantages persist.}

{\em A channel model of phase noise.---}
Light propagation in a lossy noisy media is typically modeled by a phase-covariant bosonic thermal-loss channel $\calL_{\kappa, N_B}$~\cite{Weedbrook_2012,giovannetti2014ultimate} described by the \QZ{beam-splitter transformation}
$ 
\hat{a} \to \sqrt{\kappa} \hat{a}+\sqrt{1-\kappa} \hat{e},
$ 
on the input annihilation operator $\hat{a}$,
with $\hat{e}$ in a thermal state of mean photon number $N_B/(1-\kappa)$.

\begin{figure}[t]
    \centering
    \includegraphics[width=0.475\textwidth]{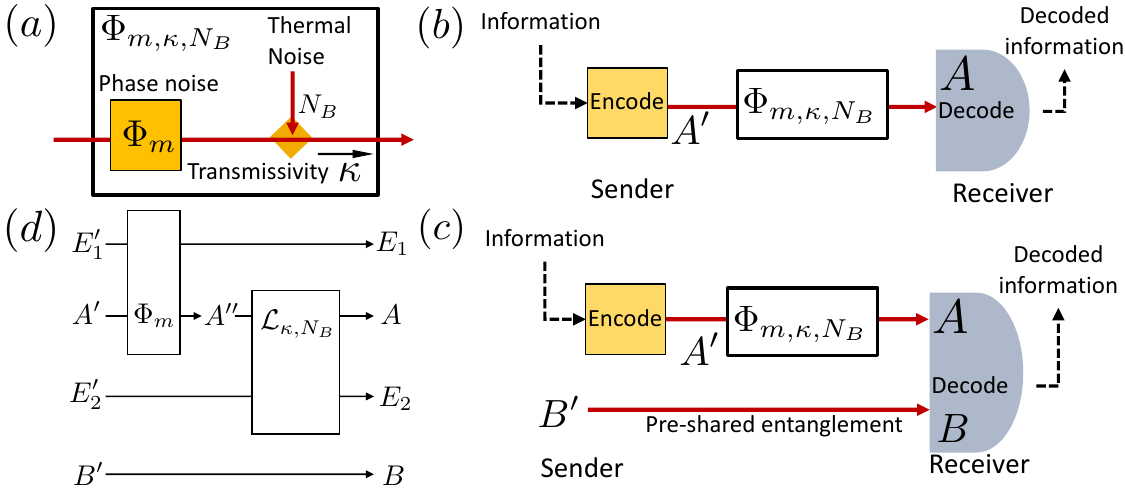}
    \caption{
    \QZ{(a) Schematic of the overall thermal-loss dephasing channel $\Phi_{m,\kappa,N_B}$, which is composed of a pure dephasing channel $\Phi_m$ and a bosonic thermal-loss channel with transmissivity $\kappa$ and noise $N_B$.
    (b) Schematic of a quantum/classical communication protocol. 
    (c) Schematic of an EA communication protocol. 
    }
    (d) Channel diagram to assist the information-theoretical analyses. Stinespring dilations are shown for both channels, with environment $E_1^\prime$ and $E_2^\prime$.
    \label{fig:schematic}
    }
\end{figure}

In additional to the noise and loss, the quantum state of light propagation also picks up a phase. In a bosonic dephasing channel $\Phi_m$ \QZ{introduced by Ref.~\cite{fanizza2020classical}}, the $m$-mode input state $\hat{\sigma}$ experiences an identical but fully random phase, leading to the output
\be 
\Phi_m(\hat{\sigma})=\braket{ \hat{U}_\theta \hat{\sigma} \hat{U}_\theta ^\dagger}_\theta\equiv \frac{1}{2\pi}\int_0^{2\pi} {d\theta}\ \hat{U}_\theta \hat{\sigma} \hat{U}_\theta ^\dagger =\sum_{n=0}^\infty p_n \hat{\sigma}_n,
\label{phi_m_def}
\ee 
where $\hat{U}_\theta=e^{i\theta \hat{n}}$ and $\hat{n}$ is the total photon number operator. The channel effectively projects the input to subspaces, each with a different fixed total photon number $n$, through the projectors $\hat{\Pi}_n=\sum_{|\bm n|=n} \ketbra{\bm n}$, where $\bm n=(n_1,\cdots, n_m)$ is an m-dimensional vector representing the photon number in each mode, and we denote $|\bm n|=\sum_{k=1}^m n_k$. Therefore, we have the final expression in Eq.~\eqref{phi_m_def} with $p_n=\tr(\hat{\Pi}_n \hat{\sigma})$ and $\hat{\sigma}_n=\hat{\Pi}_n \hat{\sigma} \hat{\Pi}_n/p_n$.
For later use, we introduce the complementary channel
$ 
\Phi_m^c (\hat{\sigma})=\sum_{k=0}^\infty \sum_{|\bm n|=k}\braket{\bm n |\hat{\sigma}|\bm n} \state{k},
$
which produces the environment $E_1$ in a Stinespring dilation (see Fig.~\ref{fig:schematic}(d)). As $\Phi_m^c$ is entanglement-breaking~\cite{horodecki2003}, $\Phi_m$ is a Hadamard channel~\cite{king2005properties} and its entire capacity region~\cite{hsieh2008entanglement,hsieh2010entanglement,wilde2012public} is additive~\cite{bradler2010trade}.

Combining the above, light propagation process can be modeled as an overall thermal-loss dephasing channel
\be 
\Phi_{m,\kappa,N_B}=\calL_{\kappa,N_B}^{\otimes m}\circ \Phi_m=\Phi_m \circ \calL_{\kappa,N_B}^{\otimes m},
\ee  
as shown in Fig.~\ref{fig:schematic}(a), where $\calL_{\kappa, N_B}^{\otimes m}$ commutes with $\Phi_m$.

{\em Capacity formula.---}
\QZ{In a general communication scenario (see Fig.~\ref{fig:schematic}(b)(c)), the sender encodes the quantum or classical information into the signal input $A^\prime$ and sends the signal through the media, e.g. a fiber or open-space link. Upon receiving the signal $A$, the receiver decodes the classical or quantum information. In an EA communication scenario, in addition to the above, the sender and receiver preshare entanglement between the input $A^\prime$ and an idler $B^\prime$, potentially through a quantum network. The idler is stored in the receiver's quantum memory for later assistance in the decoding. }
To facilitate the information-theoretical analyses, we introduce the diagram of channels in Fig.~\ref{fig:schematic}(d), where the channels are shown as Stinespring dilations acting on environment $E_1^\prime, E_2^\prime$ and the inputs. 

\QZ{
First, we start with the transmission of quantum information, without entanglement-assistance.
The ultimate rate of quantum communication over a general quantum channel $\Phi$ is given by a regularized maximization 
\be
Q(\Phi)=\lim_{N\to \infty} \frac{1}{N} \max_{\hat{\sigma}} J\left(\hat{\sigma},\Phi^{\otimes N}\right),
\label{Q_capacity}
\ee
over the input state $\hat{\sigma}$ of $A^\prime$, where the coherent information for a single channel use
$
J\left(\hat{\sigma},\Phi\right)=S\left(\Phi\left(\hat{\sigma}\right)\right)-S\left(\Phi^c\left(\hat{\sigma}\right)\right)=S\left(\hat{\rho}_A\right)-S\left(\hat{\rho}_{E_1E_2}\right)
$, as shown in Fig.~\ref{fig:schematic}(d).
Here $\hat{\rho}$ is the output state and the von Neumann entropy $S(\hat{\delta})\equiv-\tr(\hat{\delta}\log_2\hat{\delta})$ for state $\hat{\delta}$. 
As the input Hilbert space dimension is infinite, we adopt the widely-used~\cite{bennett2002entanglement,shi2020practical,fanizza2020classical,Weedbrook_2012} mean photon number constraint 
$
mE
$ 
per channel use during the entire $N\to\infty$ channel uses, such that the capacity is finite. 
}

Suppose one has pre-shared entanglement-assistance, the rate of quantum information transmission is increased to a maximization of the quantum mutual information between the received signal $A$ and the idler $B$~\cite{bennett2002entanglement}
\be 
Q_{\rm EA}(\Phi)=\frac{1}{2}C_{\rm EA}(\Phi)=\frac{1}{2}\max_{\hat{\phi}}I(A:B)_{\hat{\rho}},
\label{CE_mutual}
\ee 
over the input state $\hat{\phi}$ of $A^\prime$ and $B^\prime$. \QZ{In the first equality, we utilize the fact that the EA quantum capacity is precisely half of the EA classical capacity, via the reduction from teleportation~\cite{bennett1993} and superdense-coding~\cite{bennett1992}.}
As shown in Fig.~\ref{fig:schematic}(d), the quantum mutual information $I(A:B)_{\hat{\rho}}=S(\hat{\rho}_A)+S(\hat{\rho}_B)-S(\hat{\rho}_{E_1E_2})$, where we utilized $S(\hat{\rho}_{AB})=S(\hat{\rho}_{E_1E_2})$, due to the purity of $ABE_1E_2$. 



When a phase reference is present, the EA capacity $C_{\rm EA}(\calL_{\kappa,N_B})$ of the thermal-loss channel $\calL_{\kappa,N_B}$ is exactly solvable and known to be achieved by the TMSV encoding~\cite{bennett2002entanglement,wilde2012information,anshu2019building,qi2018applications}, thanks to the Gaussian nature of the channel; similarly, the HSW capacity~\cite{giovannetti2014ultimate} $C(\calL_{\kappa,N_B})$ can be achieved by a Gaussian ensemble of coherent states~\cite{supp}. 
\QZ{On the other hand, even with a phase reference present, only upper and lower bounds of the quantum capacity $Q\left(\calL_{\kappa,N_B}\right)$ are known~\cite{sharma2018bounding,rosati2018narrow,noh2018quantum,noh2020enhanced}; Except in the noiseless case $N_B=0$, exact solution is known and achieved by a thermal state.}

{\em Exact solution for a pure dephasing channel.---}
We begin our analyses with a pure dephasing channel in the absence of additional noise ($N_B=0, \kappa=1$). The HSW classical capacity 
$
C(\Phi_{m})/m=g(E)
$ 
can be solved~\cite{fanizza2020classical}, where $g(n)=(n+1)\log_2(n+1)-n\log_2 n$ is the entropy of a thermal state with mean photon number $n$. Extending to the \QZ{quantum-enabled case} is much more involved due to the complication from entanglement adding to the non-Gaussian nature of the problem; our first result is an exact solution of the \QZ{quantum capacity and EA capacity} of the pure dephasing channel $\Phi_m$. As $\Phi_m$ is projective in the total photon number bases \QZ{and Hadamard}, we can reduce the optimization in \QZ{Eqs.~\eqref{Q_capacity} and \eqref{CE_mutual}} to an optimization over the $m$-mode photon number distribution $P_{\bm n}$, leading to~\cite{supp}
\QZ{
\begin{subequations}
\begin{align}
&Q\left(\Phi_m\right)=\max_{\{P_{\bm n}\}} \left[H\left(\{P_{\bm n}\}_{\bm n}\right)-H\left(\{P_n^t\}_n\right)\right],
\label{Q_classical1_main}
\\
&C_{\rm EA}=2Q_{\rm EA}=\max_{\{P_{\bm n}\}} \left[2H\left(\{P_{\bm n}\}_{\bm n}\right)-H\left(\{P_n^t\}_n\right)\right],
\label{C_EA_noiseless}
\end{align}
\label{QCE}
\end{subequations}}
under the energy constraint 
$
\sum_n nP_n^t=mE
$, 
with the total photon number distribution $P_n^t=\sum_{|\bm n|=n}
P_{\bm n}$.
For $m=1$, we can immediately see that \QZ{$Q\left(\Phi_m\right)=0$ and} $C_{\rm EA}(\Phi_m)=C(\Phi_m)$ due to $P_n^t=P_{|\bm n|}$; therefore \QZ{quantum advantage in absence of phase reference is impossible for the single-mode case}. As a thermal state maximizes the von Neumann entropy given the energy constraint, we can directly obtain an upper bound as \QZ{$Q\left(\Phi_m\right)\le mg(E)$} and $C_{\rm EA}(\Phi_m)\le 2mg(E)=2C(\Phi_{m})$.

\begin{figure}[t]
    \centering
    \includegraphics[width=0.5\textwidth]{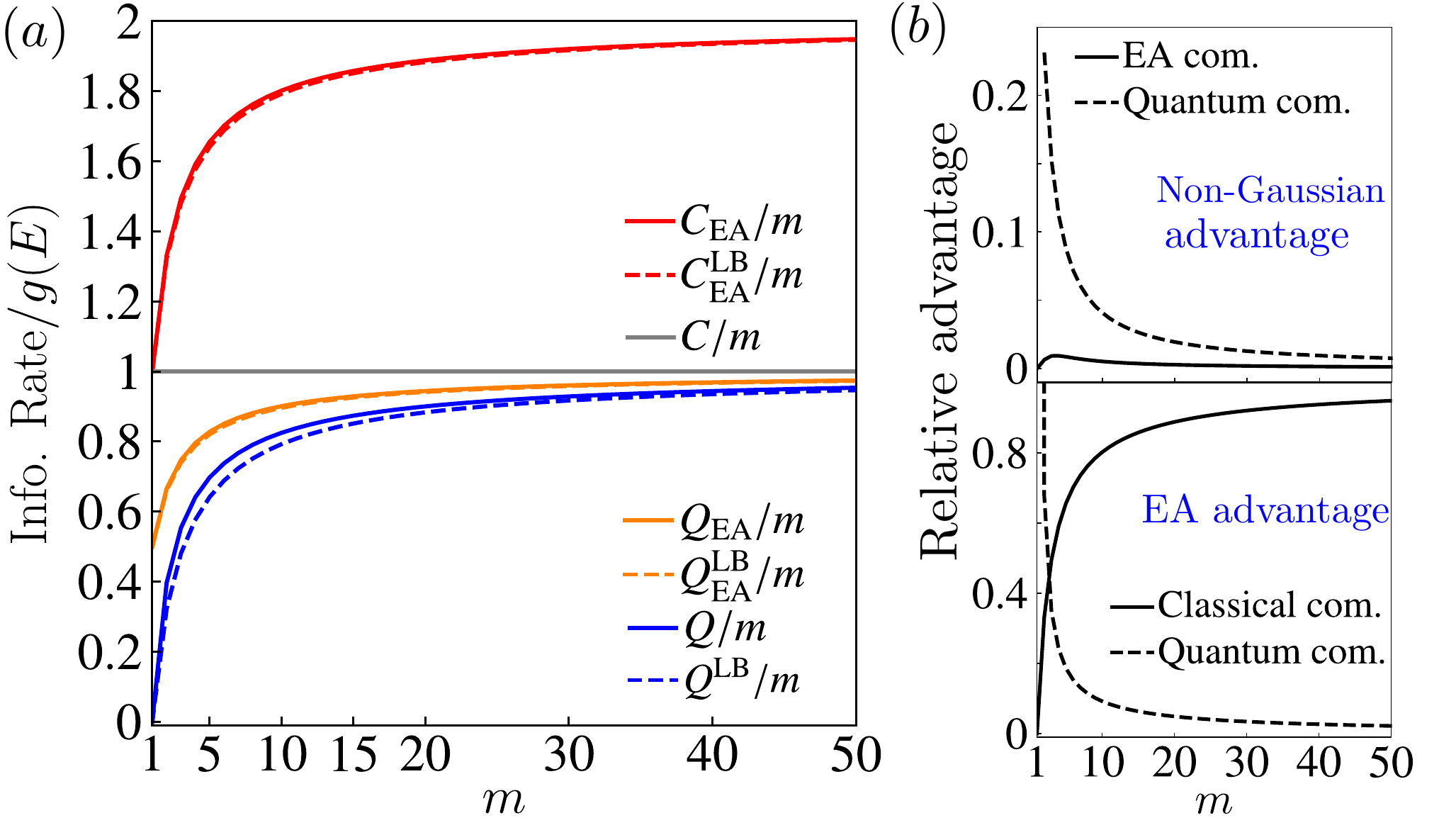}
    \caption{ 
    \QZ{(a) Ratio of the information rate over the classical capacity per mode $C=g(E)$ (gray solid) for a pure dephasing channel $\Phi_m$, with an input energy per mode $E=1$: EA classical capacity per mode $C_{\rm EA}/m$ (red solid), EA quantum capacity per mode $Q_{\rm EA}/m$ (orange solid) and quantum capacity per mode $Q/m$ (blue solid). The dashed lines with the corresponding color are the lower bound offered by Gaussian encoding schemes. 
    (b) Relative advantages in information rate for non-Gaussian encoding (up panel) and entanglement-assistance (bottom panel). 
    }
    \label{fig:CE}
    }
\end{figure}

The optimization in Eqs.~\eqref{QCE} can be solved exactly. Noticing that the constraint only affects $P_n^t$, and that $P_{\bm n}$ only appears in the first entropy term, it is optimal to distribute $P_n^t$ equally between $P_{\bm n}$ with identical $|\bm n|=n$. Then, utilizing \QZ{Lagrange} multipliers, \QZ{one obtains the optimal $m$-mode input
\be 
\hat{\sigma}_{A^\prime}=\sum_{|\bm n|=mE}\left(C_{|\bm n|+m-1}^{m-1}\right)^{-1}\state{\bm n}_{A^\prime}.
\label{optimum_input_dephasing_main}
\ee 
up to integer rounding~\cite{supp} for the quantum capacity,} and the optimal $2m$-mode input state~\cite{supp}
\be 
\hat{\phi}_{A^\prime B^\prime}\propto\sum_{\bm n}\sqrt{C_{|\bm n|+m-1}^{m-1} \tilde{\lambda}_1^{|\bm n|}}  \ket{\bm n}_{A^\prime} \ket{\bm n}_{B^\prime}
\label{phi_AB_opt_main}
\ee 
for EA communication, where $\tilde{\lambda}_1$ is determined by the energy constraint and \QZ{$C_{a}^{b}=(a!)/(b!(a-b)!)$ is the binomial coefficient}.
\QZ{The corresponding $
Q\left(\Phi_m\right)=\log_2\left(C_{mE+m-1}^{m-1}\right)
$
up to integer rounding~\cite{supp}, and $C_{\rm EA}=2Q_{\rm EA}$} can be evaluated efficiently from Eq.~\eqref{C_EA_noiseless} with the above distribution.
Interestingly, as $|\bm n|=\sum_{k=1}^m n_k$, the prefactor $C_{|\bm n|+m-1}^{m-1}$ prevents $P_{\bm n}$ to be written as any product of distributions over each variable $n_k$, making the optimal input $\hat{\phi}_{A^\prime B^\prime}$ a non-Gaussian $2m$-mode multipartite-entangled state~\cite{supp}. 

\QZ{
As we see in Fig.~\ref{fig:CE}(a), the quantum capacity (blue solid) increases from zero quickly as the memory $m$ increases; at the same time entanglement provides advantages in both quantum communication (orange solid) and classical communication (red solid), as emphasized in Fig.~\ref{fig:CE}(b) bottom panel. 
It is worthy to note that these capacity-achieving inputs provide strict advantages over sub-optimal Gaussian inputs, which are optimal in presence of a phase reference. As a consequence, the thermal input provides a lower bound for the quantum capacity $Q^{\rm LB}$, while an independent and identical (iid) product of TMSV states provides lower bounds for EA capacities $C_{\rm EA}^{\rm LB}$ and $Q_{\rm EA}^{\rm LB}$, as shown in dashed lines in Fig.~\ref{fig:CE}(a). The advantage is larger in the quantum communication case, up to $20\%$; while smaller for the EA quantum/classical communication, up to $1\%$, as shown in Fig.~\ref{fig:CE}(b) top panel.
}


{\em Capacity bounds for thermal-loss dephasing.---}
\QZ{
The exact evaluation of the capacities for $\Phi_{m,\kappa,N_B}$ in presence of loss and noise is challenging~\cite{supp}. Instead, we obtain upper and lower bounds. Combining the data-processing inequality (bottomneck inequality) and the upper bound in Refs.~\cite{noh2018quantum,sharma2018bounding}, we can obtain an upper bound
\be 
Q^{(\rm UB)}\left(\Phi_{m,\kappa,N_B}\right)\equiv \min\left[Q\left(\Phi_m\right), m \min_{1\le G_1\le 1+N_B} f(G_1) \right],
\label{QUBAll_main}
\ee 
where $Q\left(\Phi_m\right)$ is given by the exact solution and
\begin{align} 
f(G_1)=\max\big[g\left(\bar{\eta}E^\prime\right)
-g\left(\left(1-\bar{\eta}\right)E^\prime\right),0\big]
\end{align}
is derived for upper bounding $Q\left(\calL_{\kappa,N_B}\right)$. Here the constants
$E^\prime=G_2E+(G_2-1)$ 
,
$
\bar{\eta}=1-(N_B+1-\eta)/G_1
$
and
$
G_2={\eta}/[{G_1-(N_B+1-\eta)}]
$.
Similarly, for the EA capacity we have
$
C_{\rm EA}(\Phi_{m,\kappa,N_B})\le mC_{\rm EA}(\calL_{\kappa,N_B}),
$
is upper-bounded by the EA capacity of the thermal-loss channel.
}


\QZ{
On the other hand, we obtain the lower bounds
\begin{subequations}
\begin{align}
&J\left(\hat{\rho},\Phi_{m,\kappa,N_B}\right)\ge S\left(\hat{\rho}_A\right)-S\left(\hat{\rho}_{E_1}\right)-S\left(\hat{\rho}_{E_2}\right),
\label{J_sub}
\\
&I(A:B)_{\hat{\rho}}
\ge S(\hat{\rho}_A)+S(\hat{\rho}_B)-S(\hat{\rho}_{E_1})-S(\hat{\rho}_{E_2}),
\label{CE_LB}
\end{align}
\label{LB_symbols}
\end{subequations}
from the subadditivity of entropy $S(\hat{\rho}_{E_1 E_2})\le S(\hat{\rho}_{E_1})+S(\hat{\rho}_{E_2})$.
In general, there will be correlations between $E_1$ and $E_2$ and the above bounds are not tight. 
The above inequality provides ways to circumvent the non-Gaussian nature of the channel as we explained below. 
}

\QZ{
Consider Fig.~\ref{fig:schematic}(d), suppose we input a photon number diagonal state as $A^\prime$, we have $A^{\prime\prime}$ in a state identical to the input $A^\prime$; Moreover, the reduced state of the environment mode $E_2$, and that of output $A$, $B$ are identical to the states in a scenario without the channel $\Phi_m$. Now suppose the inputs are Gaussian, the only non-Gaussian part $E_1$ involved in Ineqs.~\eqref{LB_symbols} can also be efficiently calculated from the total photon number distribution. Therefore, considering iid thermal states $\hat{\rho}_{\rm th}^{\otimes m}$ for the coherent information and iid TMSVs for the EA capacity, from Ineqs.~\eqref{LB_symbols} we have the lower bounds
\begin{subequations}
\begin{align} 
&Q^{\rm LB}\equiv 
J\left(\hat{\rho}_{\rm th}^{\otimes m},\calL_{\kappa,N_B}^{\otimes m}\right)-H(\{P_n^t\}),
\\
& C_{\rm EA}^{\rm LB}\equiv 2Q_{\rm EA}^{\rm LB} \equiv  mC_{\rm EA}(\calL_{\kappa,N_B})-H(\{P_n^t\}),
\label{CEA_LB}
\end{align}
\label{LB_final}
\end{subequations}
where both the coherent information $J\left(\hat{\rho}_{\rm th}^{\otimes m},\calL_{\kappa,N_B}^{\otimes m}\right)$ and the EA capacity $C_{\rm EA}(\calL_{\kappa,N_B})$ have closed form solutions~\cite{supp}; The Shannon entropy $H(\cdot)$ is over distribution
\be 
P_n^t= C_{n+m-1}^{m-1}\frac{E^{n}}{(E+1)^{n+m}}.
\label{Pn}
\ee 
Although not having a closed form, it can be efficiently evaluated numerically. 
Furthermore, for $m\gg1$, from the law of large numbers, $P_n^t$ approaches a Gaussian distribution with mean $mE$ and variance $m E\left(E+1\right)$, therefore asymptotically
$
H(\{P_n^t\})\simeq \log_2(\epsilon \sqrt{m E\left(E+1\right)}),
$
where $\epsilon=\sqrt{2\pi e}\simeq 4.13$ is a constant. In most of the parameter region, we indeed see a good agreement between the above asymptotic expressions and exact results~\cite{supp}.  An important take-away of Ineqs.~\eqref{LB_final} is that the degradation caused by the phase noise is only of the order of $\log_2(m)/m$ bits per mode. 
}

\begin{figure}
 \centering
\includegraphics[width=0.45\textwidth]{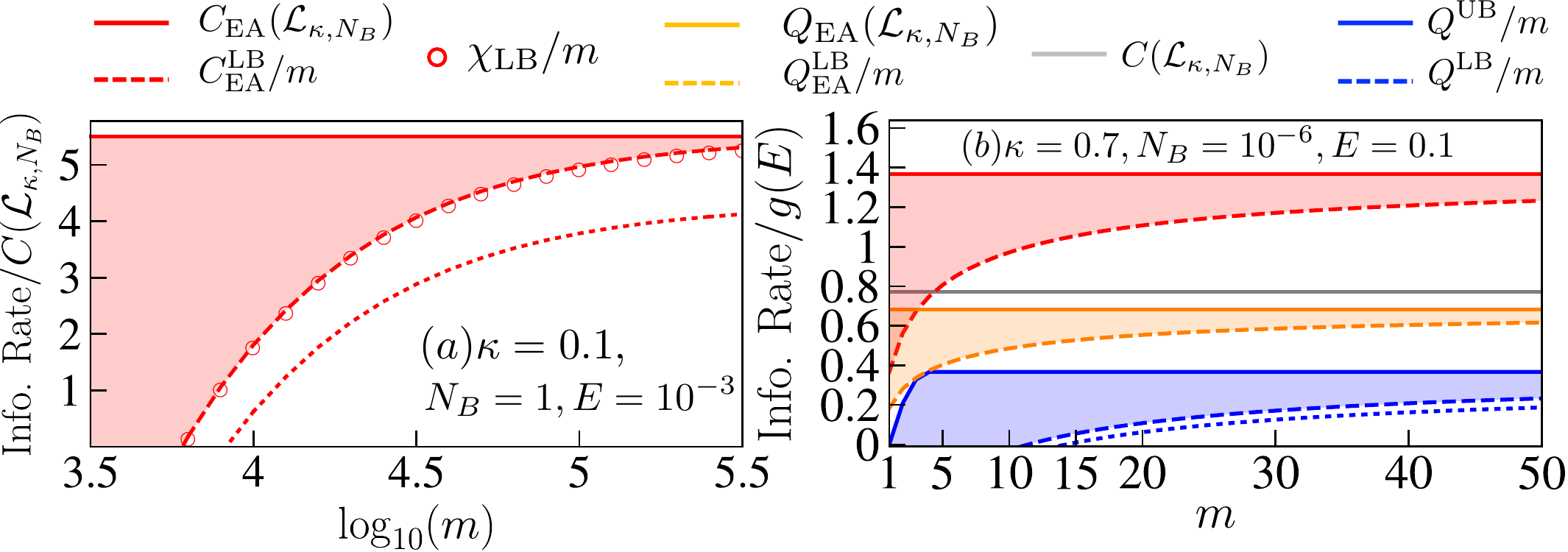}
\caption{\QZ{Information rate ratio. The EA classical capacity (red), EA quantum capacity (orange) and quantum capacity (blue) lies in the colored region. With upper bound in solid colored lines and lower bounds in dashed colored lines.
(a) Ratio of information rate over the classical capacity $C(\mathcal{L}_{\kappa,N_B})$ of a thermal-loss channel, with an input power $E=10^{-3}$ per mode and the channel transmissivity fixed at $\kappa=0.1$ and noise $N_B=1$. $\chi_{\rm LB}/m$ (red circles) is the lower bound of the accessible information of phase encoding. 
(b) Ratio of information rate over $g(E)$, with an input power $E=0.1$ per mode and $\kappa=0.7,N_B=10^{-8}$. The classical capacity of a thermal-loss channel $C(\calL_{\bar{\kappa},N_B}$ is also plotted in gray solid line for comparison. The dotted red line in (a) is the lower bound with Rayleigh fading of $\bar{\kappa}=0.1$ for EA classical capacity, and the dotted blue line in (b) is the lower bound for quantum capacity with flat fading in $\kappa\in[0.4,1]$. 
}
\label{fig:CE_noisy_new}
}
\end{figure}

\QZ{
We consider our upper and lower bounds in two typical examples: Fig.~\ref{fig:CE_noisy_new}(a) represents a microwave region where background noise $N_B\sim 1$ and transmissivity $\kappa=0.1$ (Note that larger $N_B$ provides similar results~\cite{supp}). Fig.~\ref{fig:CE_noisy_new}(b) represents an optical fiber connection or near-field free-space link, where $N_B\sim  10^{-6}$~\cite{shapiro2005ultimate}. In these plots, the upper bounds are plotted with solid lines, while lower bounds with dashed line of the same color. The colored region indicates where the true capacity lies in.
In the noisy case, we don't know the exact classical capacity of $\Phi_{m,\kappa,N_B}$, so we take the ratio of the information rate over $C(\mathcal{L}_{\kappa,N_B})\ge C(\Phi_{m,\kappa,N_B})/m$ in Fig.~\ref{fig:CE_noisy_new}(a). 
Similar to the noiseless case, the lower bound converges to the upper bound $C_{\rm EA}(\calL_{\kappa,N_B})$ quickly and revives the huge advantage over the HSW classical capacity when noise $N_B$ is large~\cite{bennett2002entanglement,shi2020practical}. The number of modes for saturation $m\sim 10^5$ in the microwave region, which corresponds to less than a \QZ{millisecond} for a \QZ{gigahertz} bandwidth microwave source.
While for the optical case of Fig.~\ref{fig:CE_noisy_new}(b), the saturation is even faster, and we see advantages both in classical communication (red vs. gray) and quantum communication (orange vs. blue) at very small $m$.
}


{\em Encoding achieving the EA lower bound.---}
With upper and lower bounds in hand, we now proceed to investigate practical encoding schemes for EA communication in presence of phase noise. \QZ{Below, we will focus on the EA classical communication; EA quantum communication can be completed with the same protocol plus teleportation.} 
We consider the performance of independent phase encoding on an iid product of TMSVs---applying a phase rotation $\hat{U}_\theta$ with uniform $\theta\in[0,2\pi)$ on each signal mode, which has been shown to saturate the EA capacity in the absence of phase noise~\cite{shi2020practical}. Combing the optimality results in Ref.~\cite{shi2020practical}, we can use the same technique in deriving the capacity lower bounds to obtain the accessible (Holevo) information~\cite{supp}
\be 
\frac{1}{m}\chi_{\rm LB}=  C_{\rm EA}(\calL_{\kappa,N_B})-\frac{1}{m}H(\{P_n^t\})+O\left(1/N_B^2\right),
\ee 
which achieves the capacity lower bound in the $N_B\gg1$ limit. We verify the above conclusions numerically in Fig.~\ref{fig:CE_noisy_new}(a). When $N_B=1$ the accessible information lower bound per mode $\chi_{\rm LB}/m$ (red open circles) overlaps with the EA capacity lower bound (red dashed).

\QZ{{\em Extension to fading channels.---}
For dynamic links such as wireless links to mobile devices~\cite{sklar1997rayleigh}, environmental fluctuations can affect more than just the phase, but also cause the transmissivity to vary with time; therefore, the overall channel output on the $m$-mode input $\hat{\sigma}$
\be 
\calR_{m,\bar{\kappa},N_B}\left(\hat{\sigma}\right) =\braket{\Phi_{m,x^2,N_B}\left(\hat{\sigma}\right)}_{f(x)},
\ee   
where $f(x)$ can be any type of fading distribution in $[0,1]$. 
We can also obtain efficiently calculable lower bounds of the capacities of $\calR_{m,\bar{\kappa},N_B}$ as~\cite{supp}
\begin{align} 
&  \frac{1}{m}C_{\rm EA}^{\rm LB}\equiv \frac{2}{m}Q_{\rm EA}^{\rm LB} \equiv \frac{1}{m}Q^{\rm LB}+g(E) \equiv \braket{g\left(x^2 E+N_B\right)}_{f(x)}
\nonumber
\\
&
+g(E)
-g\left(\nu_+\right)-g\left(\nu_-\right)-H(\{P_n^t\})/m,
\end{align}
where $\nu_\pm=\left(\sqrt{(A+S)^2-4C^2}\pm (S-A)-2\right)/4$, with the constants $A\equiv 2\left(\bar{\kappa}E+N_B\right)+1, S\equiv 2E+1$ and $C\equiv 2\braket{x}_{f(x)}\sqrt{E(E+1)}$.
}

\QZ{
For the EA communication scenario, as transmissivity is typically low, we adopt the Rayleigh distribution~\cite{goodman1976some,goodman1965some,zhuang2017fading}
$ 
f(x)\sim x\exp\left(-{x^2}/{\bar{\kappa}}\right)
$
with proper truncation in $[0,1]$; For quantum communication, the capacity is sharply zero when $\kappa<0.5$, therefore it is more reasonable to consider a flat fading of $x^2\in [\bar{\kappa}-\delta \kappa,\bar{\kappa}+\delta \kappa]$. We evaluate the lower bounds in Fig.~\ref{fig:CE_noisy_new} as the dotted lines for the same average transmissivity in each case. We see the quantum capacity is only mildly decreased due to fading (dotted blue line in Fig.~\ref{fig:CE_noisy_new}(b)); while the lower bound is decreased appreciably by fading for EA communication (dotted red line in Fig.~\ref{fig:CE_noisy_new}(a)), however, the scaling of the advantage over the HSW classical capacity survives~\cite{supp}
}

{\em Discussion.---}
In this letter, we show that \QZ{quantum advantages} in communication is possible without a phase reference, assuming a finite-time memory effect in the channel. We exactly solve the EA capacities \QZ{and quantum capacity} of a pure dephasing channel, showing that the optimal input of non-Gaussian multipartite-entangled state is strictly better than the Gaussian entangled source. In presence of additional thermal noise, loss \QZ{and fading effects}, we derive upper and lower bounds of the capacities, which shows that the degradation from the absence of a phase reference is mild. Many future directions can be explored, including an extension to cases with finite phase noise~\cite{arqand2020quantum} and practical protocol design for achieving the lower bounds.

Q.Z. acknowledges the Defense Advanced Research Projects Agency (DARPA) under Young Faculty Award (YFA) Grant No. N660012014029 and Craig M. Berge Dean's Faculty Fellowship of University of Arizona.

\appendix

\tableofcontents

\section{Full capacity formula for the thermal-loss channel} 
Under the input energy constraint $E$, the EA classical capacity for a thermal loss channel $\calL_{\kappa,N_B}$ is solved in Ref.~\cite{bennett2002entanglement}
\be
C_{\rm EA}(\calL_{\kappa,N_B})=g(E)+g(E^\prime)-g(A_+)-g(A_-),
\label{CE_formula}
\ee 
where $g(n)=(n+1)\log_2(n+1)-n\log_2 n$ is the entropy of a thermal state with energy $n$, and \ba
&&A_\pm=(D-1\pm(E^\prime-E))/2, 
\\
&&E^\prime=\kappa E+N_B 
\\
&&D=\sqrt{(E+E^\prime+1)^2-4\kappa E(E+1)}.
\ea
Moreover, it is also known that its HSW capacity~\cite{giovannetti2014ultimate} under the energy constraint $E$ can be achieved by a Gaussian ensemble of coherent states, giving
\be 
C(\calL_{\kappa,N_B})=g(\kappa E+N_B)-g(N_B).
\label{C_formula}
\ee 

\section{Exact solution for a pure dephasing channel}

\subsection{EA capacity}

In the noiseless case only a single environment $E_1$ is needed.
Consider the most general input state 
\be 
\hat{\phi}_{A^\prime B^\prime}=\sum_{\bm n}\sqrt{P_{\bm n}}  \ket{\bm n}_{A^\prime} \ket{\chi_{\bm n}}_{B^\prime},
\label{phi_AB}
\ee 
where in general the states $\ket{\chi_{\bm n}}$ do not have to be an orthogonal basis. The output state can be written as
\begin{align}
	&\hat{\rho}_{AB}= {\Phi_m}_{\ A^\prime \to A} \otimes \calI_{B^\prime \to B} (\hat{\phi}_{A^\prime B^\prime})
	\nonumber
	\\
	&=\sum_{n=0}^\infty
	\sum_{|\bm n|=n}\sum_{|\bm n^\prime|=n}
	\sqrt{P_{\bm n}P_{\bm n^\prime}}  \ket{\bm n}_{A}\bra{\bm n^\prime}\otimes  \ket{\chi_{\bm n}}_{B}    \bra{\chi_{\bm n^\prime}}. 
\end{align}
One can obtain the environment from the complementary channel
\be 
\hat{\rho}_{E_1}=\Phi_m^c (\hat{\rho})=\sum_{n=0}^\infty P_n^t \state{n},
\ee 
where we introduced $P_n^t=\sum_{|\bm n|=n} P_{\bm n}$.
Therefore, the entropy
\be 
S(\hat{\rho}_{E_1})=H\left(\{P_n^t\}_n\right),
\label{SE1}
\ee 
is the entropy of the distribution of total photon number. 

Because initially $\hat{\phi}_{A^\prime B^\prime}$ is pure and $\calI_{B^\prime \to B}$ is identity
\be 
S(\hat{\rho}_B) = S(\hat{\phi}_{B^\prime})=
S\left(\sum_{\bm n}
P_{\bm n}  \state{\chi_{\bm n}}_{B} \right)\le H\left(\{P_{\bm n} \}_{\bm n}\right),
\label{SB}
\ee 
where equality is achieved when $\ket{\chi_{\bm n}}$ are orthonormal bases. Here $H(\cdot)$ denotes the Shannon entropy over a classical distribution.

For $S(\hat{\rho}_A)$, we can upper bound it by introducing the fully dephasing channel on each of the $m$ modes. 
Because a fully de-phasing channel $\Phi_1^{\otimes m}$ is effectively a photon number (projective) measurement channel, $\Phi_1^{\otimes m}$ only increases the entropy of the input quantum state, we have
\begin{align} 
	&S(\hat{\rho}_A)\le S\left(\Phi_1^{\otimes m}\left(\hat{\rho}_A\right)\right)
	\nonumber
	\\
	&=
	S\left(\sum_{\bm n}
	P_{\bm n}  \ket{\bm n}_{A}    \bra{\bm n}\right)=H\left(\{P_{\bm n}\}_{\bm n}\right),
	\label{SA}
\end{align} 
where the equality can be achieved by choosing $\{\ket{\chi_{\bm n}}\}_{\bm n}$ as an orthonormal bases. 

Combining Eqs.~\eqref{SE1},~\eqref{SB} and~\eqref{SA}, we have from Eq.~\eqref{CE_mutual} of the main paper 
\begin{align}
	C_{\rm EA}(\Phi_m)\le\max_{\hat{\phi}}
	\left[2H\left(\{P_{\bm n} \}_{\bm n}\right)
	-H\left(\{P_n^t\}_n\right)\right],
\end{align}
where the inequality can be achieved by choosing $\{\ket{\chi_{\bm n}}\}_{\bm n}$ as orthonormal bases, e.g.
\be 
\ket{\chi_{\bm n}}=\ket{\bm n}.
\ee 
Under this optimum choice, the mutual information only depends on the distribution $P_{\bm n}$, and
\be
C_{\rm EA}(\Phi_m)=\max_{P_{\bm n}}\left[ 2H\left(\{P_{\bm n}\}_{\bm n}\right)-H\left(\{P_n^t\}_n\right)\right]
\label{CE_max}
\ee  
under the energy constraint, $\tr(\hat{n}\hat{\phi}_A)=mE$, or
\be 
\sum_{n=0}^\infty P_n^t n=mE.
\label{P_E_constraint}
\ee 
This optimization can be solved exactly. First, we fix the total photon number distribution $P_n^t$ and solve the optimal $m$-mode photon number distribution $P_{\bm n}$. Noticing that the constraint Eq.~\eqref{P_E_constraint} only affects $P_n^t$, and $P_{\bm n}$ only appears in the first entropy term in Eq.~\eqref{CE_max}, it is straightforward that for a fixed $P_n^t$, it is optimal to have
\be 
P_{\bm n}=\frac{P_{|\bm n|}^t}{C_{|\bm n|+m-1}^{m-1}},
\ee 
equal for all of the $C_{|\bm n|+m-1}^{m-1}$ photon number patterns with equal total photon number $|\bm n|$. Here $C_{a}^{b}=(a!)/(b!(a-b)!)$ is the binomial coefficient. Then Eq.~\eqref{CE_max} is simplified to
\begin{align}
	&C_{\rm EA}(\Phi_m)=\max_{P_n^t} F, \mbox{ with }
	\\
	& F=\sum_{n=0}^\infty \left[-2{P_{n}^t}\log_2\left(\frac{P_{n}^t}{C_{n+m-1}^{m-1}}\right)+P_n^t\log_2\left(P_n^t\right)\right]
	\nonumber
	\\
	&=\sum_{n=0}^\infty \left[-{P_{n}^t}\log_2\left(\frac{P_{n}^t}{\left(C_{n+m-1}^{m-1}\right)^2}\right)\right].
	\label{CE_max_2}
\end{align}
Now we solve the constrained optimization by introducing Lagrange multipliers
\be 
F_{\lambda_1,\lambda_2}=F-\lambda_1\left(\sum_{n=0}^\infty P_n^t n-mE\right)-\lambda_2\left(\sum_{n=0}^\infty P_n^t-1\right).
\ee 
The optimal condition
\be 
\partial_{P_n^t}F_{\lambda_1,\lambda_2}=-\log_2\left(\frac{P_{n}^t}{\left(C_{n+m-1}^{m-1}\right)^2}\right)-\frac{1}{\ln(2)}-\lambda_1 n-\lambda_2=0
\ee 
leads to the solution
\be 
P_{n}^t=\left(C_{n+m-1}^{m-1}\right)^2 \exp\left(-\lambda_1 n-\lambda_2\right),
\ee 
where we absorbed the constant $1$ into $\lambda_2$.
By the normalization condition and energy constraint Eq.~\eqref{P_E_constraint}, we have
\ba
&&e^{-\lambda_2}{}_2F_1(m,m,1,e^{-\lambda_1})=1,
\\
&&e^{-\lambda_1-\lambda_2}m^2 {}_2F_1(m+1,m+1,2,e^{-\lambda_1})=mE,
\ea
where ${}_2F_1$ is the hypergeometric function,
leading to the final expression
\be
P_{n}^t=\frac{\left(C_{n+m-1}^{m-1}\right)^2}{{}_2F_1(m,m,1,\tilde{\lambda}_1)} \tilde{\lambda}_1^n,
\ee 
where $\tilde{\lambda}_1$ is the solution to
\be
\frac{\tilde{\lambda}_1 m\ {}_2F_1(m+1,m+1,2,\tilde{\lambda}_1)}{{}_2F_1(m,m,1,\tilde{\lambda}_1)}= E.
\ee 
The corresponding EA classical capacity $C_{\rm EA}(\Phi_m)$ can be evaluated through Eq.~\eqref{CE_max_2}.

The optimal state achieving the performance has the form
\be 
\hat{\phi}_{A^\prime B^\prime}=\sum_{\bm n}\sqrt{P_{\bm n}}  \ket{\bm n}_{A^\prime} \ket{\bm n}_{B^\prime},
\label{phi_AB_opt}
\ee 
with 
\be 
P_{\bm n}=\frac{\left(C_{|\bm n|+m-1}^{m-1}\right)}{{}_2F_1(m,m,1,\tilde{\lambda}_1)} \tilde{\lambda}_1^{|\bm n|}.
\ee 
Interestingly, as $|\bm n|=\sum_{k=1}^m n_k$, $P_{\bm n}$ cannot be written as a product of distributions over each variable $n_k$. As the input $\hat{\phi}_{A^\prime B^\prime}$ is pure, this means that the optimum input is a multipartite entangled state. 

It is easy to see such a state is non-Gaussian. For example, for the $m=2$ case, the reduced distribution
\begin{align}
&P_{n_1}=\sum_{n_2=0}^\infty P_{\bm n=(n_1,n_2)}
=\frac{1-\tilde{\lambda}_1}{1+\tilde{\lambda}_1}\tilde{\lambda}_1^{n_1}\left(\left(1-\tilde{\lambda}_1\right)n_1+1\right).
\end{align}
The only photon-number diagonal Gaussian state is a thermal state. We see the above is not a thermal distribution, therefore the overall state cannot be Gaussian.

\subsection{Quantum capacity}
\label{sec:exact_solution}
We begin our analyses with a pure dephasing channel in the absence of additional noise ($N_B=0, \kappa=1$). In this case, as the channel is Hadamard, we can restrict the optimization to a single-letter as
\be 
Q_{mE}\left(\Phi_m\right)=\max_{\hat{\rho}} J(\hat{\rho}, \Phi_m)
\ee 
under the energy constraint of $mE$. In the supplemental material, we will make the energy constraint (e.g. $mE$ here) explicit, as it is necessary in later part of detailed analyses. Making use of the channel and the complementary channel
\begin{align}
J(\hat{\rho}, \Phi_m)&=S\left(\Phi_m\left(\hat{\rho}\right)\right)-S\left(\Phi^c_m\left(\hat{\rho}\right)\right)
\\
&\le S\left(\Phi_1^{\otimes m}\circ \Phi_m\left(\hat{\rho}\right)\right)-S\left(\Phi^c_m\left(\hat{\rho}\right)\right)
\\
&= S\left(\sum_{\bm n} P_{\bm n}\state{\bm n}\right)-S\left(\sum_{n=0}^\infty P_n^t \state{n}\right)
\\
&=H\left(\{P_{\bm n}\}_{\bm n}\right)-H\left(\{P_n^t\}_n\right)
\end{align} 
where the inequality is achieved with photon number diagonal inputs and we denote $P_{\bm n}=\braket{\bm n|\hat{\rho}|\bm n}$ and $P_n^t=\sum_{|\bm n|=n}P_{\bm n}$. In the last line, we have reduced the von Neumann entropy to Shannon entropy over probability distributions. 
Therefore the quantum capacity is now a maximization over classical probability distribution
\be 
Q_{mE}\left(\Phi_m\right)=\max_{\{P_{\bm n}\}} \left[H\left(\{P_{\bm n}\}_{\bm n}\right)-H\left(\{P_n^t\}_n\right)\right].
\label{Q_classical1}
\ee 
And the energy constraint is
\be 
\sum_{n=0}^\infty P_n^t n=m E.
\ee 

First, we notice that for a fixed choice of $P_n^t$, the optimum choice of $P_{\bm n}$ is to have the most symmetric distribution
\be 
P_{\bm n}=\frac{P_{|\bm n|}^t}{C_{|\bm n|+m-1}^{m-1}},
\ee 
such that the first term in Eq.~\eqref{Q_classical1} is maximized and then we can simplify the formula
\begin{align}
Q_{mE}\left(\Phi_m\right)&=\max_{\{P_n^t\}}
\left[\sum_{n=0}^\infty {P_{n}^t}\log_2\left(C_{n+m-1}^{m-1}\right)-H\left(\{P_n^t\}_n\right)\right].
\end{align} 
Because $\log_2\left(C_{n+m-1}^{m-1}\right)$ as a function of $n$ is concave, we have
\be 
\sum_{n=0}^\infty {P_{n}^t}\log_2\left(C_{n+m-1}^{m-1}\right)\le 
 \log_2\left(C_{m E+m-1}^{m-1}\right).
\ee 
Therefore, the maximum is achieved when the distribution is centered at a single $n=mE$. However, now we have to take care of the fact that $n$ is an integer. To satisfy this, one can consider the $N\to \infty$ limit, and adopt the time sharing strategy of sending states with either $\lfloor m E \rfloor$ or $\lfloor m E \rfloor+1$ photons to saturate the mean photon number constraint, and thus the probability of of sending each state is: 
\begin{align}
&p_1=\lfloor m E \rfloor+1-m E,
\\
&p_2=m E-\lfloor m E \rfloor,
\end{align}
Therefore the precise quantum capacity is
\begin{align}
&Q_{mE}\left(\Phi_m\right)=
(\lfloor m E \rfloor+1-m E)\log_2\left(C_{\lfloor m E \rfloor+m-1}^{m-1}\right)+
\nonumber
\\
&(m E-\lfloor m E \rfloor)\log_2\left(C_{\lfloor m E \rfloor+m}^{m-1}\right).
\end{align}
we denote the above as 
\be 
Q_{mE}\left(\Phi_m\right)=\lfloor\log_2\left(C_{mE+m-1}^{m-1}\right)\rfloor.
\label{QPhim}
\ee 
Therefore the optimum input is a photon number diagonal state with a fixed total photon number $mE$, and equal distribution over all patterns of photon number with $|\bm n|=mE$, i.e.,
\be 
\hat{\rho}=\sum_{|\bm n|=mE}\frac{1}{C_{|\bm n|+m-1}^{m-1}}\state{\bm n}.
\label{optimum_input_dephasing}
\ee 
up to subtleties from integer rounding procedures.

The limit of $m\to \infty$ while $E$ being a constant is interesting to look at, as one expects it to approach the noiseless case. Indeed, we find
\be 
\lim_{m\to \infty} \frac{1}{m} Q_{mE}\left(\Phi_m\right)=g(E),
\ee 
where $g(n)=(n+1)\log_2(n+1)-n\log_2 n$ is the entropy of a thermal state with mean photon number $n$.

We can also calculate the rate from iid thermal states, which are optimal for the scenario without dephasing. The total photon number distribution of $m$ iid thermal states
\be 
P_n^t= C_{n+m-1}^{m-1}\frac{E^{n}}{(E+1)^{n+m}}.
\label{Pn}
\ee 
Then we have 
\begin{align}
J(\hat{\rho}_{\rm th}^{\otimes m}, \Phi_m)=mg(E)-H(\{P_n^t\})
\\
\simeq mg(E)-\log_2(\epsilon \sqrt{m E\left(E+1\right)}),
\end{align}
where $\epsilon=\sqrt{2\pi e}\simeq 4.13$ is a constant.

\section{Challenges in the exact solution}

Consider the same input in Eq.~\eqref{phi_AB}. Ineq.~\eqref{SA} can be extended as
\begin{align}
	&S(\hat{\rho}_A)\le S\left(\Phi_1^{\otimes m}\circ \Phi_{m,\kappa,N_B}\left(\hat{\rho}_A\right)\right)
	\nonumber
	\\
	&=S\left(\Phi_1^{\otimes m}\circ \Phi_m \circ \calL_{\kappa,N_B}^{\otimes m}\left(\hat{\rho}_A\right)\right)
	\nonumber
	\\
	&=S\left(\Phi_1^{\otimes m}\circ \calL_{\kappa,N_B}^{\otimes m}\left(\hat{\rho}_A\right)\right)
	\nonumber
	\\
	&=
	S\left(\calL_{\kappa,N_B}^{\otimes m}\left(\sum_{\bm n}
	P_{\bm n}  \ket{\bm n}_{A}    \bra{\bm n}\right)\right),
\end{align} 
where in the last step we utilized the covariant nature of $\calL_{\kappa,N_B}$. Similar to Ineq.~\eqref{SA}, the inequality can be achieved by choosing $\{\ket{\chi_{\bm n}}\}_{\bm n}$ as an orthonormal bases.

The output state can be written as
\begin{align}
	&\hat{\rho}_{AB}=\left(\calL_{\kappa,N_B}^{\otimes m} \circ {\Phi_m} \right)_{A^{\prime}\to A} \otimes \calI_{B^\prime \to B} (\hat{\phi}_{A^\prime B^\prime})
	\\
	&=\sum_{n=0}^\infty
	\sum_{|\bm n|=|\bm n^\prime|=n}
	\sqrt{P_{\bm n}P_{\bm n^\prime}}  \calL_{\kappa,N_B}^{\otimes m}\left(\ket{\bm n}_{A}\bra{\bm n^\prime}\right)\otimes  \ket{\chi_{\bm n}}_{B}    \bra{\chi_{\bm n^\prime}}. 
\end{align}
Due to the channel $\calL_{\kappa,N_B}^{\otimes m}$, it is now challenging to further simplify the expression. Similarly, the complementary channel now involves two correlated non-Gaussian ancilla $E_1$ and $E_2$ and the further simplification becomes difficult.

\section{Extension to lossy noisy case}

\subsection{EA capacity}

\subsubsection{Upper bounds}
To begin with, we obtain an upper bound of $C_{\rm EA}$, as the quantum mutual information satisfies the data-processing inequality, we have
\begin{align} 
&\frac{1}{m}C_{\rm EA}(\Phi_{m,\kappa,N_B})=\frac{1}{m}C_{\rm EA}\left(\Phi_{m}\circ \calL_{\kappa,N_B}^{\otimes m}\right)
\nonumber
\\
&\le \frac{1}{m}C_{\rm EA}\left( \calL_{\kappa,N_B}^{\otimes m}\right)= C_{\rm EA}(\calL_{\kappa,N_B}),
\label{CE_UB}
\end{align}
is upper-bounded by the EA capacity of the thermal-loss channel,
where in the last step we utilized the additivity of the EA capacity.

\subsubsection{Lower bounds}
With the upper bound in hand, we now obtain a lower bound through the subadditivity of entropy, $S(\hat{\rho}_{E_1 E_2})\le S(\hat{\rho}_{E_1})+S(\hat{\rho}_{E_2})$, therefore Eq.~\eqref{CE_mutual} leads to
\be
C_{\rm EA}
\ge \max_{\hat{\phi}}\left[S(\hat{\rho}_A)+S(\hat{\rho}_B)-S(\hat{\rho}_{E_1})-S(\hat{\rho}_{E_2})\right].
\label{CE_LB}
\ee
In general, there will be correlations between $E_1$ and $E_2$ and the above bound is not tight. We consider a product of TMSVs as the input $\hat{\phi}$ to further obtain a lower bound. 
Although the joint state $\hat{\rho}_{E_1E_2}$ is still non-Gaussian, in the following we show that the reduced states $\hat{\rho}_A$, $\hat{\rho}_B$ and $\hat{\rho}_{E_2}$ are all Gaussian, which enables the entropy evaluation. Furthermore, the entropy of $\rho_{E_1}$ can be efficiently calculated, despite being non-Gaussian.

The state of the environment $E_1$ can be calculated from the complementary channel $\Phi_m^c$ as a photon number diagonal state
$
\hat{\rho}_{E_1}=\sum_{n=0}^\infty P_n^t \state{n},
$ 
where the total photon number distribution of $m$ iid thermal states is given by Eq.~\eqref{Pn}.
Therefore, its von Neumann entropy $S\left(\hat{\rho}_{E_1}\right)=H(\{P_n^t\})$ reduces to classical Shannon entropy; although not having a closed form, it can be efficiently evaluated numerically. 
Furthermore, for $m\gg1$, from the law of large numbers, $P_n^t$ approaches a Gaussian distribution with mean $mE$ and variance $m E\left(E+1\right)$, therefore asymptotically
\be 
S\left(\hat{\rho}_{E_1}\right)=H(\{P_n^t\})\simeq \log_2(\epsilon \sqrt{m E\left(E+1\right)}),
\label{Hasym}
\ee 
where $\epsilon=\sqrt{2\pi e}\simeq 4.13$ is a constant.

Now we consider the environment $E_2$. As the state of the input $A^\prime$ is photon number diagonal, the reduced state of the output $A^{\prime \prime}$ of $\Phi_m$ is identical to the input state of an iid product of thermal states $\hat{\phi}_{A^\prime}$. 
Thus, the reduced state of the environment mode $\hat{\rho}_{E_2}=\calL_{\kappa,N_B}^{c\otimes m} (\hat{\phi}_{A^\prime})$ is identical to the state in a scenario without the channel $\Phi_m$. Moreover, the reduced state of the final output $\hat{\rho}_A=\calL_{\kappa,N_B}^{\otimes m}(\hat{\phi}_{A^\prime})$ is again identical to that without the channel $\Phi_m$. The same applies to $\hat{\rho}_B$. Combing the above analyses, Ineq.~\eqref{CE_LB} can be further lower bounded as
\be 
C_{\rm EA}\ge S(\calL_{\kappa,N_B}^{\otimes m}(\phi_{A^\prime}))+S(\hat{\rho}_B)-S(\calL_{\kappa,N_B}^{c\otimes m} (\phi_{A^\prime}))-H(\{P_n^t\}).
\ee 
Noticing that the first three terms give the EA capacity $C_{\rm EA}(\calL_{\kappa,N_B}^{\otimes m})$, we have the lower bound
\be
C_{\rm EA}/m\ge C_{\rm EA}^{\rm LB}/m\equiv C_{\rm EA}(\calL_{\kappa,N_B})-H(\{P_n^t\})/m
\label{CEA_LB}
\ee
where we utilized the additivity of the EA classical capacity. We can also make use of the asymptotic expression in Eq.~\eqref{Hasym} to obtain the asymptotic lower bound 
\be 
\frac{1}{m}C_{\rm EA}^{\rm LB, asym} \equiv C_{\rm EA}(\calL_{\kappa,N_B})-\frac{1}{m}\log_2(\epsilon \sqrt{m E\left(E+1\right)}).
\label{CEA_LB_asym}
\ee
When $m$ goes to infinity, the upper bound in Eq.~\eqref{CE_UB} and the above lower bound coincide and we have $C_{\rm EA}/m= C_{\rm EA}(\calL_{\kappa,N_B})$, which converges to the case with a phase reference present. The degradation caused by the phase noise is only of the order of $\log_2(m)/m$ per mode.

\subsection{Quantum capacity}

Although the pure-dephasing case can be solved exactly, in general the quantum capacity of $\Phi_{m,\kappa,N_B}$ is challenging to compute. As in general the regularization in Eq.~\eqref{Q_capacity} is necessary; the non-Gaussian nature of the channel further complicates the analyses, while even for the thermal loss channel the exact solution of the quantum capacity is also an open question. In this section, we will focus on obtaining upper and lower bounds of the quantum capacity.


\subsubsection{Upper bounds}
We begin with the upper bound.
Because of bottleneck inequality, we have
\be 
Q_{mE}\left(\Phi_{m,\kappa,N_B}\right)\le \min\left[Q_{mE}\left(\calL_{\kappa,N_B}^{\otimes m}\right),Q_{mE}\left(\Phi_m\right) \right].
\label{QUBAll}
\ee 
We have the exact solution for $Q_{mE}\left(\Phi_m\right)$ in Eq.~\eqref{QPhim}, while for $Q_{mE}\left(\calL_{\kappa,N_B}^{\otimes m}\right)$ we can further obtain an upper bound from data-processing. Consider the channel decomposition relation $\calL_{\eta,N_B}=\calA_{G_1}\circ \calL_{\bar{\eta},0}\circ \calA_{G_2}$, with
\begin{align}
\bar{\eta}=1-\frac{1}{G_1}(N_B+1-\eta)
\\
G_2=\frac{\eta}{G_1-(N_B+1-\eta)}.
\end{align} 
Again due to bottleneck inequality, we can use the energy constrained quantum capacity of the pure loss channel $\calL_{\bar{\eta},0}$ to upper bound the quantum capacity as~\cite{noh2018quantum,sharma2018bounding}
\be 
Q_{E^\prime}\left(\calL_{\kappa,N_B}\right)\le  Q_{E^\prime}^{(\rm UB)}\left(\calL_{\kappa,N_B}\right)\equiv \min_{1\le G_1\le 1+N_B} f_{\eta, N_B,E^\prime}(G_1),
\ee 
where the function
\begin{align}
f_{\eta, N_B,E^\prime}(G_1)=&\max \big[g\left(\bar{\eta}\left(G_2E^\prime+(G_2-1)\right)\right)
\nonumber
\\
&-g\left(\left(1-\bar{\eta}\right)\left(G_2E^\prime+\left(G_2-1\right)\right)\right),0\big].
\end{align} 
No that our definition of $N_B$ is the amount of noise mixed in, and therefore the formula look different.
One can check that $f_{\eta, N_B,E^\prime}(G_1)$ is a concave function in $E^\prime$.
Note that the energy-constrained upper bound is tight when $N_B=0$. It is also easy to check that in the infinite energy limit, $Q_{E^\prime}^{(\rm UB)}\left(\calL_{\kappa,N_B}\right)$ converges to the energy-unconstrained bound in Refs.~\cite{sharma2018bounding,rosati2018narrow,noh2018quantum},
\be 
\lim_{E\to\infty} Q_{E^\prime}^{(\rm UB)}\left(\calL_{\kappa,N_B}\right)=  \max \left[\log_2\left(\frac{\eta-N_B}{N_B+1-\eta}\right),0\right].
\ee

To obtain the upper bound for the channel $\calL_{\eta,N_B}^{\otimes m}$, one needs to deal with the energy constraint, which is $mE$ total mean photon number across all $m$ modes, therefore it is not a trivial problem. We consider the overall channel decomposition
\be
\calL_{\eta,N_B}^{\otimes m}=\otimes_{k=1}^m \calA_{G_1^k}\circ \calL_{\bar{\eta}^k,0}\circ \calA_{G_2^k},
\ee
where the coefficients $\{G_1^k\}$, $\{G_2^k\}$ and $\{\bar{\eta}^k\}$ are not necessarily equal for different $k$. Then we apply the bottleneck inequality
\begin{align}
&Q_{mE}\left(\calL_{\kappa,N_B}^{\otimes m}\right)
\\
&\le \min_{\{G_1^k\}} Q_{mE_t}\left(\otimes_{k=1}^m\calL_{\bar{\eta}^k,0}\right)  
\label{UBS1}
\\
&\le \min_{\{G_1^k\}}  \sum_{k}  Q_{E_t^k}\left(\calL_{\bar{\eta}^k,0}\right) 
\label{UBS2}
\\
&= \min_{\{G_1^k\}} \sum_{k} f_{\eta, N_B,E^k}(G_1^k)
\label{UBS3}
\\
&\le m\min_{G} \frac{1}{m}\sum_{k} f_{\eta, N_B,E^k}(G)
\label{UBS4}
\\ 
&\le m\min_{G}  f_{\eta, N_B,\frac{1}{m}\sum_{k} E^k}(G)
\label{UBS5}
\\
&= m\min_{G}  f_{\eta, N_B,E}(G)=m Q_{E}^{(\rm UB)}\left(\calL_{\kappa,N_B}\right).
\label{UBS6}
\end{align}
In \eqref{UBS1}, $mE_t$ is the total energy after the first amplification $\calA_{G_2^k}$ on each inputs, when the total input energy is constrained by $mE$. In \eqref{UBS2}, we utilized the additivity of pure-loss channel, and the energy constrain $E_t^k$ is determined by each of the amplifier $G_2^k$ and input energy. In \eqref{UBS3}, we utilized the capacity formula for pure-loss channel. In \eqref{UBS4}, we fixed all gains to be identical $G$, which will increase the minimization value. In \eqref{UBS5}, we utilized the concavity of the function $f_{\eta, N_B,E^\prime}(G_1)$ in $E^\prime$. In \eqref{UBS6}, we utilized the energy constraint $\sum_{k=1}^m E^k=mE$.

Combining \eqref{UBS6} and \eqref{QPhim}, Ineq.~\eqref{QUBAll} leads to
\begin{align}
&\frac{1}{m}Q_{mE}\left(\Phi_{m,\kappa,N_B}\right)\le \frac{1}{m}Q_{mE}^{(\rm UB)}\left(\Phi_{m,\kappa,N_B}\right)
\\
&\equiv \min\left[ Q_{E}^{(\rm UB)}\left(\calL_{\kappa,N_B}\right),\frac{1}{m}\lfloor\log_2\left(C_{mE+m-1}^{m-1}\right)\rfloor \right].
\label{QUB_final}
\end{align}

\subsubsection{Lower bounds}
Now we obtain the lower bound. We consider the input of a product of thermal state to lower bound the capacity. The non-Gaussian nature of the channel complicates the problem and we adopt the following inequality to circumvent the non-Gaussian nature of the channel. From Fig.~\ref{fig:schematic} (b), the coherent information of the overall channel can be written as
\begin{align}
J\left(\hat{\rho},\Phi_{m,\kappa,N_B}\right)&=S\left(\hat{\rho}_A\right)-S\left(\hat{\rho}_{E_1E_2}\right)
\\
&\ge S\left(\hat{\rho}_A\right)-S\left(\hat{\rho}_{E_1}\right)-S\left(\hat{\rho}_{E_2}\right),
\end{align}
where we utilized the subadditivity of entropy.
Suppose we input photon number diagonal state as $A^\prime$, we have $A^{\prime\prime}$ in a state identical to the input $A^\prime$; therefore if we have a lower bound on quantum capacity of $\calL_{\kappa,N_B}$ achieved by photon number diagonal state we can obtain a lower bound for $\Phi_{m,\kappa,N_B}$.

To begin with, we can consider iid thermal states and obtain
\begin{align} 
&J\left(\hat{\rho}_{\rm th}^{\otimes m},\Phi_{m,\kappa,N_B}\right)\ge
\nonumber
\\
&Q_{mE}^{\rm LB}\left(\Phi_{m,\kappa,N_B}\right)\equiv 
J\left(\hat{\rho}_{\rm th}^{\otimes m},\calL_{\kappa,N_B}^{\otimes m}\right)-H(\{P_n^t\}),
\label{Q_LB}
\end{align}
where the original lower bound is
\begin{align}
\frac{1}{m}J\left(\hat{\rho}_{\rm th}^{\otimes m},\calL_{\kappa,N_B}^{\otimes m}\right)
&=g\left(\kappa E+N_B \right)
\nonumber
\\
&-g\left(\frac{D+\left(1-\kappa\right)E-N_B-1}{2}\right)
\nonumber
\\
&-g\left(\frac{D-\left(1-\kappa\right)E+N_B-1}{2}\right),
\end{align}
with $D=\sqrt{\left(\left(1+\kappa\right)E+N_B+1\right)^2-4\kappa E\left(E+1\right)}$.

\section{Detailed derivation of the phase encoding}
\label{app:detailed_de}

Here all operations commute, therefore we can effectively write the output state
\be 
\hat{\rho}_{AB}^{\bm \theta}=\left[(\otimes_{k=1}^m \hat{U}_{\theta_k})\otimes \hat{I}\right] \hat{\rho}_{AB}\left[(\otimes_{k=1}^m \hat{U}_{\theta_k})\otimes \hat{I}\right]^\dagger
\ee
conditioned on the overall phase encoding $\bm \theta=(\theta_1,\cdots,\theta_m)$ over $m$ mode pairs.

Denote $\Sigma_{\bm \theta}$ as the ensemble of $\hat{\rho}_{AB}^{\bm \theta}$ with uniform random phases.
The accessible (Holevo) information at the receiver side
\be 
\chi\left(\Sigma_{\bm \theta}\right)=S(\braket{ \hat{\rho}_{AB}^{\bm \theta}}_{\bm \theta})-\braket{ S( \hat{\rho}_{AB}^{\bm \theta})}_{\bm \theta}
\ee 
is the information rate achievable by the encoding and optimum receivers.

The conditional entropy $S( \hat{\rho}_{AB}^{\bm \theta})=S\left(\hat{\rho}_{AB}\right)$ due to unitarity of the encoding, and therefore $\braket{ S( \hat{\rho}_{AB}^{\bm \theta})}_{\bm \theta}=S\left(\hat{\rho}_{AB}\right)=S\left(\hat{\rho}_{E_1E_2}\right)$. While the average state can be produced equivalently by the fully dephasing channel $\Phi_1^{\otimes m}$ as the following   
\begin{align}
&\braket{ \hat{\rho}_{AB}^{\bm \theta}}_{\bm \theta}=\left[\Phi_1^{\otimes m} \circ \Phi_m \circ \calL_{\kappa,N_B}^{\otimes m}\right]_{A^\prime\to A} \otimes \calI_{B^\prime\to B} (\hat{\phi}_{A^\prime B^\prime})
\nonumber
\\
&=\left[\Phi_1^{\otimes m}  \circ \calL_{\kappa,N_B}^{\otimes m}\right]_{A^\prime\to A} \otimes \calI_{B^\prime\to B} (\hat{\phi}_{A^\prime B^\prime})=\braket{ \hat{\tilde{\rho}}_{AB}^{\bm \theta}}_{\bm \theta},
\label{chi_uncond}
\end{align}
which is equal to the average state $\braket{ \hat{\tilde{\rho}}_{AB}^{\bm \theta}}_{\bm \theta}$ produced when the dephasing channel $\Phi_m$ is absent.

We can use the same technique when we calculate the capacity lower bound of the main paper, as detailed in the following,
\begin{align}
	\chi\left(\Sigma_{\bm \theta}\right)&=S\left(\braket{ \hat{\tilde{\rho}}_{AB}^{\bm \theta}}_{\bm \theta}\right)-S\left(\hat{\rho}_{E_1E_2}\right)
	\label{S1}
	\\
	&\ge  S\left(\braket{ \hat{\tilde{\rho}}_{AB}^{\bm \theta}}_{\bm \theta}\right)-S\left(\hat{\rho}_{E_1}\right)-S\left(\hat{\rho}_{E_2}\right)
	\label{S2}
	\\
	&= S\left(\braket{ \hat{\tilde{\rho}}_{AB}^{\bm \theta}}_{\bm \theta}\right)-  H(\{P_n^t\}) -S\left(\hat{\tilde{\rho}}_{E_2}\right)
	\label{S3}
	\\
	&= \left[S\left(\braket{ \hat{\tilde{\rho}}_{AB}^{\bm \theta}}_{\bm \theta}\right)-\braket{S\left(\hat{\tilde{\rho}}_{AB}\right)}_{\bm \theta}\right]-  H(\{P_n^t\})
	\label{S4}
	\\
	&= \chi\left(\tilde{\Sigma}_{\bm \theta}\right)-  H(\{P_n^t\})
	\label{S5}
	\\
	&= m\chi\left(\tilde{\Sigma}_{\theta}\right)-  H(\{P_n^t\})\equiv \chi_{\rm LB}\left(\Sigma_{\bm \theta}\right),
	\label{chi_phase_LB_detailed}
\end{align}
where in Eq.~\eqref{S1} we utilized Eq.~\eqref{chi_uncond}, and the fact that $\braket{ S( \hat{\rho}_{AB}^{\bm \theta})}_{\bm \theta}=S\left(\hat{\rho}_{AB}\right)=S\left(\hat{\rho}_{E_1E_2}\right)$. In Ineq.~\eqref{S2}, we applied subadditivity of von Neumann entropy. In Eq.~\eqref{S3}, we utilized Eq.~\eqref{Hasym} of the main paper and the reduced state of the environment mode $\hat{\rho}_{E_2}=\calL_{\kappa,N_B}^{c\otimes m} (\hat{\phi}_{A^\prime})$ is identical to $\hat{\tilde{\rho}}_{E_2}$ without the channel $\Phi_m$. Eq.~\eqref{S4} is due to the purity of $E_2AB$ in absence of channel $\Phi_m$. In Eq.~\eqref{S5}, $\tilde{\Sigma}_{\bm \theta}$ is the ensemble of states produced when phase noise is absent. In the last step, $\chi\left(\tilde{\Sigma}_{\bm \theta}\right)= m\chi\left(\tilde{\Sigma}_{\theta}\right)$ due to the iid structure of the state. The quantity $\chi\left(\tilde{\Sigma}_{\theta}\right)$ has been calculated in Ref.~\cite{shi2020practical}.

We can also use the asymptotic expression in Eq.~\eqref{Hasym} to obtain the corresponding asymptotic expression
\be 
\chi_{\rm LB,asym}\left(\Sigma_{\bm \theta}\right)
\equiv  m\chi\left(\tilde{\Sigma}_{\theta}\right)-  \log_2(\epsilon\sqrt{m E\left(E+1\right)}).
\ee   
In Ref.~\cite{shi2020practical} we showed analytically that $\chi\left(\tilde{\Sigma}_{\theta}\right)=C_{\rm EA}(\calL_{\kappa,N_B})+O\left(1/N_B^2\right)$, therefore
\be 
\frac{1}{m}\chi_{\rm LB}\left(\Sigma_{\bm \theta}\right)=  C_{\rm EA}(\calL_{\kappa,N_B})-\frac{1}{m}H(\{P_n^t\})+O\left(1/N_B^2\right),
\ee 
achieves the capacity lower bound in the $N_B\gg1$ limit, which leads to the logarithmically diverging advantage over the HSW classical capacity.

\section{Extension to fading channel}


For dynamic links such as wireless links to mobile devices~\cite{sklar1997rayleigh}, environmental fluctuations can affect more than just the phase, but also cause the transmissivity to vary from time to time; therefore, the overall channel output on the $m$-mode input $\hat{\sigma}$ can be written as
\begin{align}
&\calR_{m,\bar{\kappa},N_B}\left(\hat{\sigma}\right) =\braket{\Phi_{m,x^2,N_B}\left(\hat{\sigma}\right)}_{f(x)} 
\\
&= \Phi_m \circ \overline{\calL}_{\bar{\kappa},N_B}=  \overline{\calL}_{\bar{\kappa},N_B}  \circ \Phi_m,
\end{align}   
with the ensemble averaged channel
\be 
\overline{\calL}_{\bar{\kappa},N_B}(\hat{\sigma})=\braket{\calL_{x^2,N_B}^{\otimes m}\left(\hat{\sigma}\right) }_{f(x)}.
\ee 

Below, we show that the HSW capacity 
\be 
C\left(\calR_{m,\bar{\kappa},N_B}\right)/m\le C\left(\calL_{\bar{\kappa},N_B}\right)
\ee  
is upper bounded by that of the average channel; While combining the convexity property~\cite{adami1997neumann,bennett2002entanglement,elkouss2016nonconvexity}, we have the upper bound
\be  
 C_{\rm EA}^{\rm UB}\equiv 2 Q_{\rm EA}^{\rm UB}\equiv m\braket{C_{\rm EA}(\calL_{x^2,N_B})}_{f(x)}.
\ee  
For quantum capacity, due to non-convex~\cite{smith2008quantum}, a similar bound is not obtained.
We will also provide efficiently calculable lower bounds on the capacities.

\subsection{EA capacity}
Here the ensemble average is over the Rayleigh-fading distribution~\cite{goodman1976some,goodman1965some,zhuang2017fading}
\be 
f(x)=\frac{2x}{(1-e^{-1/\tilde{\kappa}})\tilde{\kappa}} \exp\left(-\frac{x^2}{\tilde{\kappa}}\right), x\in [0,1],
\ee 
with the value $\tilde{\kappa}$ determined from the expectation value
\begin{align}
&\bar{\kappa}=\braket{x^2}_{f(x)}=\tilde{\kappa}+\frac{1}{1-e^{1/\tilde{\kappa}}}.
\end{align}
While the mean
\be 
s_1=\braket{x}_{f(x)}=\frac{\sqrt{\pi \tilde{\kappa} }{\rm Erf}\left(1/\sqrt{\tilde{\kappa}}\right)/2-e^{-1/\tilde{\kappa}}}{1-e^{-1/\tilde{\kappa}}}\simeq \sqrt{\pi \bar{\kappa}}/2.
\ee 
It is worthy to note that our approach does not rely on the particular Rayleigh type and will apply to any type of fading distribution. We are considering the fast fading scenario, where the memory effect is within a finite time period where $m$ modes can be transmitted. In the so-called slow fading case where memory effect lasts forever, it goes to the so-called compound channel that has been extensively studied in both the HSW capacity~\cite{datta2007coding,datta2009classical} and EA classical capacity~\cite{boche2017entanglement,boche2016entanglement,berta2017entanglement}.

\begin{figure}[t]
    \centering
    \includegraphics[width=0.3\textwidth]{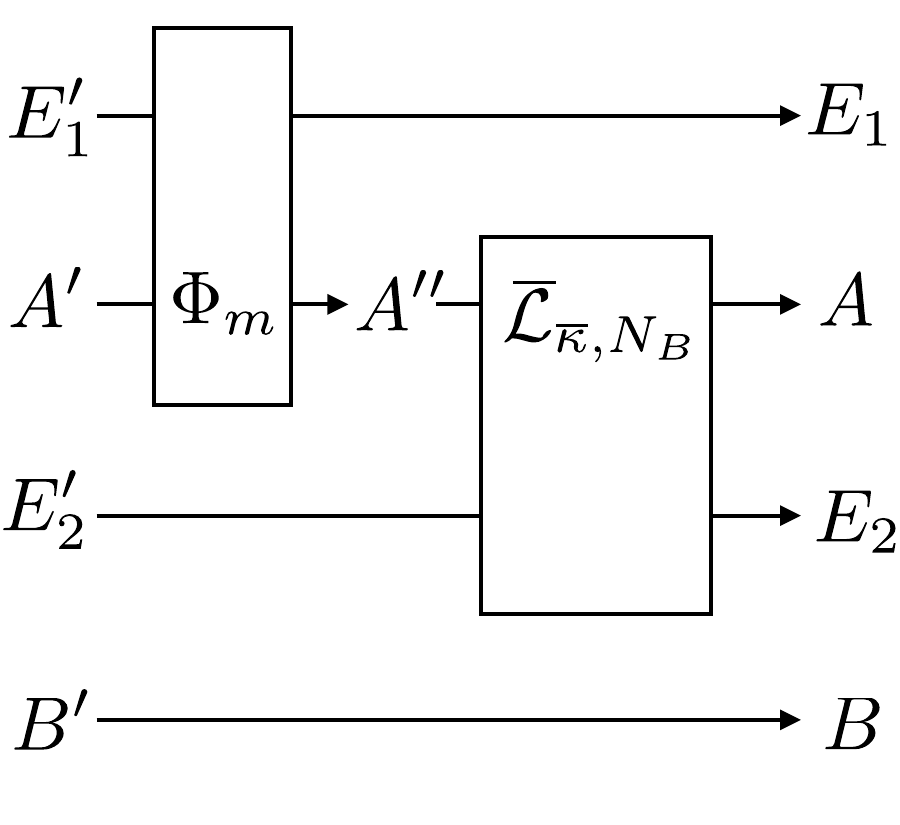}
    \caption{ Channel diagram to assist the information-theoretical analyses. Stinespring dilations are shown for both channels, with environment $E_1^\prime$ and $E_2^\prime$.
    \label{fig:schematic_fading}
    }
\end{figure}

As the memory effect is within a finite time, we can use the capacity formula for memoryless channels
\begin{align} 
&C_{\rm EA}\left(\calR_{m,\bar{\kappa},N_B}\right)
\nonumber
\\
&=\max_{\hat{\sigma}}
\left[
S(\hat{\sigma})
+S\left(\calR_{m,\bar{\kappa},N_B}\left(\hat{\sigma}\right)\right)
-S\left(\calR^c_{m,\bar{\kappa},N_B} \left(\hat{\sigma}\right)\right)
\right].
\label{CE_mutual_fading}
\end{align}
To facilitate the analysis, we can also introduce the channel diagram in Fig.~\ref{fig:schematic_fading}. This allows an alternative way of expression the formula
\begin{align} 
&C_{\rm EA}\left(\calR_{m,\bar{\kappa},N_B}\right)
=\max_{\hat{\sigma}}
\left[
S(\hat{\rho}_B)
+S\left(\hat{\rho}_A\right)
-S\left(\hat{\rho}_{E_1E_2}\right)
\right].
\label{CE_mutual_fading_ABE}
\end{align}

While the exact solution is challenging, we can obtain lower and upper bounds of the capacity.

\subsubsection{Upper bound}
First, as the EA classical capacity is convex~\cite{adami1997neumann,bennett2002entanglement,elkouss2016nonconvexity}, we have the upper bound
\begin{align} 
&C_{\rm EA}\left(\calR_{m,\bar{\kappa},N_B}\right)
\le
\braket{C_{\rm EA}\left(\Phi_{m,x^2,N_B}\left(\hat{\sigma}\right)\right)}_{f(x)}.
\end{align}
Then we can apply the upper bound Eq.~\eqref{CE_UB} and obtain the final upper bound
\be 
\frac{1}{m}C_{\rm EA}\left(\calR_{m,\bar{\kappa},N_B}\right)\le \frac{1}{m}C_{\rm EA}^{\rm UB}\equiv \braket{C_{\rm EA}(\calL_{x^2,N_B})}_{f(x)}.
\ee

\subsubsection{Lower bound}
To enable a lower bound, we specify the input $\hat{\phi}$ to be a product of TMSV state with mean photon number $E$ per mode. The reduced input state $\hat{\sigma}$ is an iid product of thermal state with mean photon number $E$ per mode, leading to the solution of the first term of Eq.~\eqref{CE_mutual_fading} as
\be 
S(\hat{\sigma})=mg(E).
\label{LBpart1}
\ee 

{\em First non-trivial term: concavity.---} 
The first non-trivial term of Eq.~\eqref{CE_mutual_fading}  is
\begin{align}
&S\left(\calR_{m,\bar{\kappa},N_B}\left(\hat{\sigma}\right)\right)
\\
&=S\left(\braket{\Phi_{m,x^2,N_B}\left(\hat{\sigma}\right)}_{f(x)} \right)
\\
&\ge 
\braket{S\left(\Phi_{m,x^2,N_B}\left(\hat{\sigma}\right) \right)}_{f(x)}
\end{align}
due to concavity of entropy. As $\Phi_m$ does not change photon number diagonal inputs, including an iid product of thermal state,
\begin{align}
&\braket{S\left(\Phi_{m,x^2,N_B}\left(\hat{\sigma}\right) \right)}_{f(x)}
\\
&
=\braket{S\left(\calL_{x^2,N_B}^{\otimes m}\circ \Phi_m\left(\hat{\sigma}\right) \right)}_{f(x)}
\\
&
=\braket{S\left(\calL_{x^2,N_B}^{\otimes m}\left(\hat{\sigma}\right) \right)}_{f(x)}
\\
&
=m \braket{g\left(x^2 E+N_B\right)}_{f(x)}
\label{LBpart2}
\end{align}
which can be numerically evaluated easily.

{\em Second non-trivial term: Gaussian extremality.---}
Now we focus on the second non-trivial term of Eq.~\eqref{CE_mutual_fading}
\begin{align}
&S\left(\calR^c_{m,\bar{\kappa},N_B} \left(\hat{\sigma}\right)\right)=S(\hat{\rho}_{E_1E_2})
\label{SS1}
\\
&\le S(\hat{\rho}_{E_1})+S(\hat{\rho}_{E_2})
\label{SS2}
\\
&= S\left(\Phi_m^c (\hat{\sigma})\right)+S\left(\overline{\calL}^c_{\bar{\kappa},N_B}\left(\Phi_m\left(\hat{\sigma}\right)\right)\right)
\label{SS3}
\\
&= S\left(\Phi_m^c (\hat{\sigma})\right)+S\left(\overline{\calL}^c_{\bar{\kappa},N_B}\left(\hat{\sigma}\right)\right)
\label{SS4}
\\
&= H(\{P_n^t\})+S\left(\hat{\zeta}_m\right),
\label{SS5}
\end{align}
where the state
\be 
\hat{\zeta}_m\equiv \overline{\calL}_{\bar{\kappa},N_B}\otimes \calI \left(\hat{\phi}_{\rm TMSV}^{\otimes m}\right)=\braket{\calL_{x^2,N_B}^{\otimes m}\otimes \calI\left(\hat{\phi}_{\rm TMSV}^{\otimes m}\right) }_{f(x)}.
\ee 
is a $2m$-mode non-Gaussian state. In step \eqref{SS1}, we utilized the alternative expression enabled by the diagram in Fig.~\ref{fig:schematic_fading}. In step \eqref{SS2}, subadditivity of von Neumann entropy is applied. Step \eqref{SS4} made use of the fact that $\Phi_m$ preserves photon number diagonal states. In the final step, an equivalent state is introduced to calculate the entropy.

We can therefore upper bound the entropy by the entropy of the Gaussian state $\hat{\zeta}_{m,G}$ with the same covariance matrix, due to Gaussian extremality~\cite{holevo1999capacity,wolf2006}, i.e.,
\be
S\left(\hat{\zeta}_m\right)\le S\left(\hat{\zeta}_{m,G}\right)
\le m S\left(\hat{\zeta}_{1,G}\right) 
\ee 
where in the last step we utilized subadditivity of von Neumann entropy again and the reduced 2-mode Gaussian state 
\be 
\hat{\zeta}_{1,G}=\braket{\calL_{x^2,N_B}\otimes \calI \left(\hat{\phi}_{\rm TMSV}\right) }_{f(x)}.
\ee 
The von Neumann entropy of the above state can be analytically calculated, due to the Gaussian nature of the state (as detailed in Sec.~\ref{cov_mat}), therefore leading to the final bound from Ineq.~\eqref{SS5}
\be 
S\left(\calR^c_{m,\bar{\kappa},N_B} \left(\hat{\sigma}\right)\right)\le H(\{P_n^t\})+m S\left(\hat{\zeta}_{1,G}\right). 
\label{LBpart3}
\ee 

{\em Overall lower bound.---}
Combining Eq.~\eqref{LBpart1}, Ineq.~\eqref{LBpart2} and Ineq.~\eqref{LBpart3} into Eq.~\eqref{CE_mutual_fading}, we have the final lower bound
\begin{align} 
&\frac{1}{m}C_{\rm EA}\left(\calR_{m,\bar{\kappa},N_B}\right)\ge 
\nonumber
\\
&
g(E)
+\braket{g\left(x^2 E+N_B\right)}_{f(x)}
- S\left(\hat{\zeta}_{1,G}\right)-H(\{P_n^t\})/m.
\end{align}

\subsubsection{Classical benchmark}
The HSW classical capacity of the channel $\calR_{m,\bar{\kappa},N_B}$ is still an open problem. To benchmark the advantage from entanglement, we need to understand the HSW capacity. As the exact solution is challenging, we will obtain an upper bound instead. To begin with, we separate out the dephasing part via the data-processing inequality
\begin{align}
C\left(\calR_{m,\bar{\kappa},N_B}\right)
&= 
C\left(\Phi_m \circ \braket{\calL_{x^2,N_B}^{\otimes m} }_{f(x)}\right)
\\
&
\le 
C\left(\braket{\calL_{x^2,N_B}^{\otimes m} }_{f(x)}\right).
\label{C_UB_all}
\end{align}
In general, it is unkown if the classical capacity is convex in the quantum channels, but here we can still upper bound the classical capacity through the following lemma (see Section~\ref{proof_lemma_supp} for a proof).

\begin{lemma}
\label{lemma:CUB}
Consider a set of channels $\{\calR_k\}_{k=1}^N$ from input state acting on Hilbert space $\calH_I$ to the output state acting on Hilbert $\calH_O$. Denote the minimum output entropy
\be 
S_{\rm min}(\calR_k)=\lim_{n\to \infty} \frac{1}{n}\min_{\hat{\rho}\in \calH_I^{\otimes n}} S(\calR_k^{\otimes n}(\hat{\rho})).
\ee 
Suppose the minimum output entropy of an arbitrary tensor product of the above channels factors i.e., $S_{\rm min}(\calR_{k_1}\otimes \calR_{k_2}\cdots )=S_{\rm min}(\calR_{k_1})+S_{\rm min}(\calR_{k_2})+\cdots$ for any $k_1,k_2,\cdots$, then the classical capacity of the average channel $\sum_k q_k \calR_k$ can be upper bounded by 
\be 
C\left(\sum_k q_k \calR_k\right)\le S_{\rm max}-\sum_k q_k S_{\rm min}(\calR_k),
\label{C_UB_fading}
\ee 
where $S_{\rm max}$ is the maximum possible entropy of quantum states acting on $\calH_O$. For $\calH_O=\mathbb{C}^d$, $S_{\rm max}=\log_2(d)$; for m-mode infinite-dimensional $\calH_O$ with energy constraint $E^\prime$ per mode, we have $S_{\rm max}=mg(E^\prime)$.

\end{lemma}

Now we look at Ineq.~\eqref{C_UB_all} again. The channel $\braket{
\calL_{x^2,N_B}^{\otimes m} }_{f(x)}$ is a convex mixture of tensored single-mode bosonic thermal-loss channels $\{\calL_{x^2,N_B}^{\otimes m}\}$. We know from Refs.~\cite{giovannetti2015solution,giovannetti2015majorization} that their minimum output entropy factors and therefore Lemma~\ref{lemma:CUB} applies. To utilize Ineq.~\eqref{C_UB_fading}, we first calculate the output energy constraint
\begin{align}
E^\prime= 
\braket{x^2 E+N_B }_{f(x)} = \bar{\kappa} E+N_B.
\end{align}
The minimum output entropy of each channel $\calL_{x^2,N_B}^{\otimes m}$ is achieved by an vacuum input and
\be 
S_{\rm min}(\calL_{x^2,N_B}^{\otimes m})=mg(N_B)
\ee 
is equal for all channels, therefore Ineq.~\eqref{C_UB_fading} gives
\be 
C\left(\braket{\calL_{x^2,N_B}^{\otimes m} }_{f(x)}\right)\le m g\left(\bar{\kappa} E+N_B\right)-m g\left(N_B\right).
\ee 
Finally, combining Ineq.~\eqref{C_UB_all}, we have
\be 
\frac{1}{m}C\left(\calR_{m,\bar{\kappa},N_B}\right)\le  g\left(\bar{\kappa} E+N_B\right)- g\left(N_B\right)\equiv C\left(\calL_{\bar{\kappa},N_B}\right).
\ee 

\subsection{Quantum capacity}

The original fading model concerns the case of low transmissivity and has a Rayleigh distribution; however, when transmissivity is below $1/2$, the quantum capacity will be zero. So here, instead of the Rayleigh fading where quantum capacity is hardly non-zero, we consider the fading effects to model the uncertainty in the transmissivity. For simplicity, we assume symmetric uniform deviation of the power transmission ratio $x^2\in[\bar{\kappa}-\delta \kappa,\bar{\kappa}+\delta \kappa]$, which corresponds to 
\be 
f(x)=\frac{x}{\delta \kappa}, x\in [\sqrt{\bar{\kappa}-\delta \kappa},\sqrt{\bar{\kappa}+\delta \kappa}].
\ee 

The overall channel output on the $m$-mode input $\hat{\sigma}$ can be written as
\begin{align}
&\calR_{m,\bar{\kappa},N_B}\left(\hat{\sigma}\right) =\braket{\Phi_{m,x^2,N_B}\left(\hat{\sigma}\right)}_{f(x)} 
\\
&= \Phi_m \circ \overline{\calL}_{\bar{\kappa},N_B}\left(\hat{\sigma}\right)=  \overline{\calL}_{\bar{\kappa},N_B}  \circ \Phi_m\left(\hat{\sigma}\right),
\end{align}   
with the ensemble averaged channel
\be 
\overline{\calL}_{\bar{\kappa},N_B}(\hat{\sigma})=\braket{\calL_{x^2,N_B}^{\otimes m}\left(\hat{\sigma}\right) }_{f(x)}.
\ee

Similar to the case without fading, we can have the bottleneck inequality
\be 
Q_{mE}\left(\calR_{m,\bar{\kappa},N_B}\right)\le \min\left[Q_{mE}\left(\overline{\calL}_{\bar{\kappa},N_B}\right),Q_{mE}\left(\Phi_m\right) \right].
\label{QUBAll_fading}
\ee 
However, upper bounding $Q_{mE}\left(\overline{\calL}_{\bar{\kappa},N_B}\right)$ is challenging, as quantum capacity is non-convex~\cite{smith2008quantum}, in contrast to the EA classical capacity~\cite{bennett2002entanglement}. The upper bound from $Q_{mE}\left(\Phi_m\right)$ still holds, however, untight when $m$ is large. 

We can obtain a lower bound on the quantum capacity, via an encoding of iid thermal state, similar to the EA classical communication case
\begin{align} 
&\frac{1}{m}Q_{mE}\left(\calR_{m,\bar{\kappa},N_B}\right)\ge  \frac{1}{m}Q_{mE}^{\rm LB}\equiv 
\braket{g\left(x^2 E+N_B\right)}_{f(x)}
\nonumber
\\
&
- g\left(\left({\mu_+-1}\right)/{2}\right)-g\left(\left({\mu_--1}\right)/{2}\right)-H(\{P_n^t\})/m,
\label{Q_LB_fading}
\end{align}
where $\mu_\pm=\left(\sqrt{(A+S)^2-4C^2}\pm (S-A)\right)/2$, with the constants $A\equiv 2\left(\bar{\kappa}E+N_B\right)+1, S\equiv 2E+1$ and $C\equiv 2\braket{x}_{f(x)}\sqrt{E(E+1)}$.

\section{Additional plots}
In this section, we provide some additional plots to support the conclusions in the main paper. 

\begin{figure*}
 \centering
\includegraphics[width=0.7\textwidth]{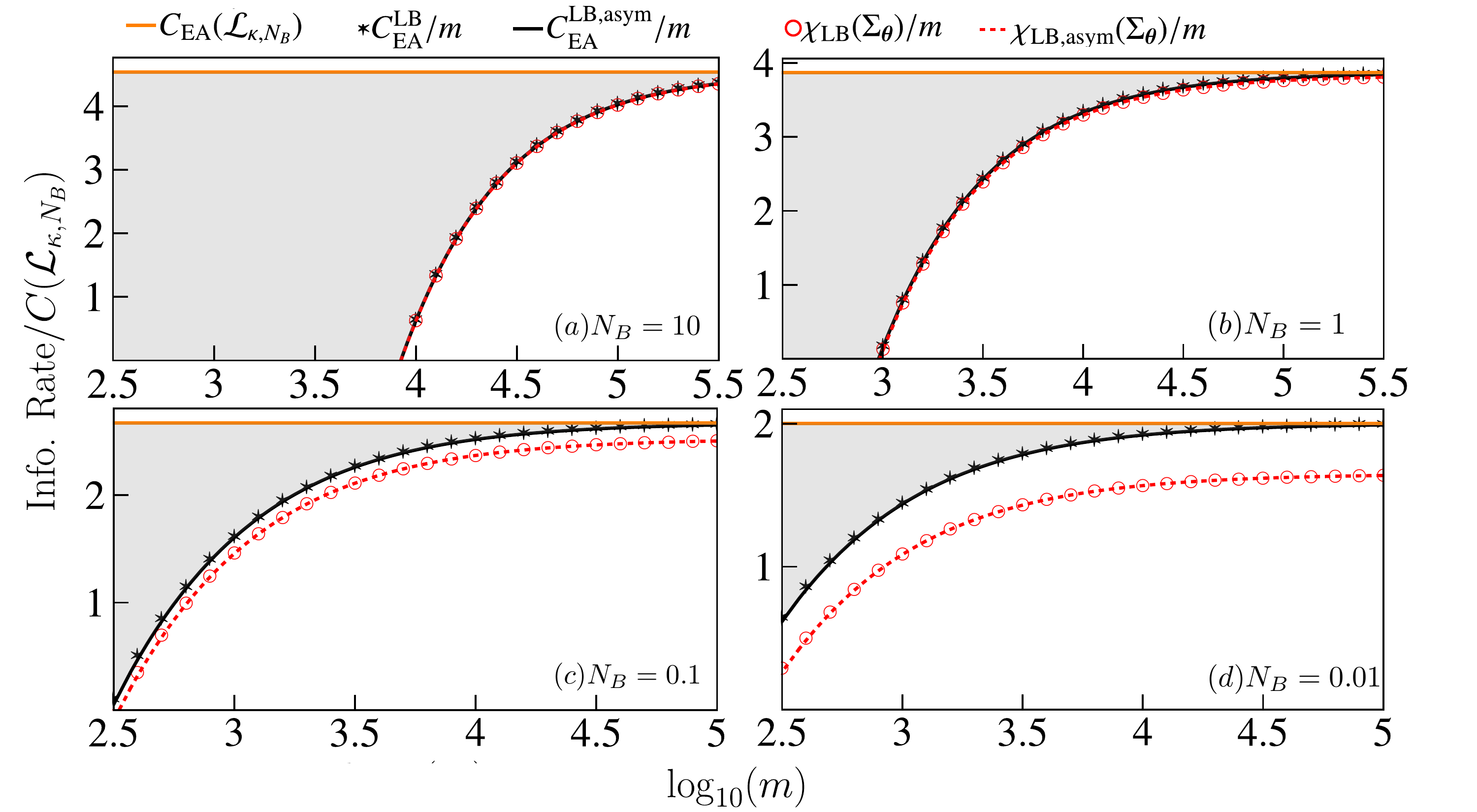}
\caption{Ratio of information rate of a thermal-loss dephasing channel $\Phi_{m,\kappa,N_B}$ over the classical capacity $C(\mathcal{L}_{\kappa,N_B})$ of a thermal-loss channel, with an input power $E=0.01$ per mode and the channel transmissivity fixed at $\kappa=0.1$. The noise $N_B=10,1,0.1,0.01$ in (a)-(d). The EA capacity per mode $C_{\rm EA}(\Phi_{m,\kappa,N_B})/m$ lies in the gray region, between the upper bound $C_{\rm EA}(\calL_{\kappa,N_B})$ (orange) and the lower bound $C_{\rm EA}^{\rm LB}/m$ (black star). The asymptotic lower bound $C_{\rm EA}^{\rm LB,asym}/m$ is plotted as the black solid line for comparison. Similarly, $\chi_{\rm LB}(\Sigma_{\bm \theta})/m$ (red circles) is the lower bound of the accessible information of phase encoding, while its asymptotic is given by $\chi_{\rm LB,asym}(\Sigma_{\bm \theta})/m$ (red dashed lines).  
\label{fig:CE_noisy}
}
\centering
\includegraphics[width=0.7\textwidth]{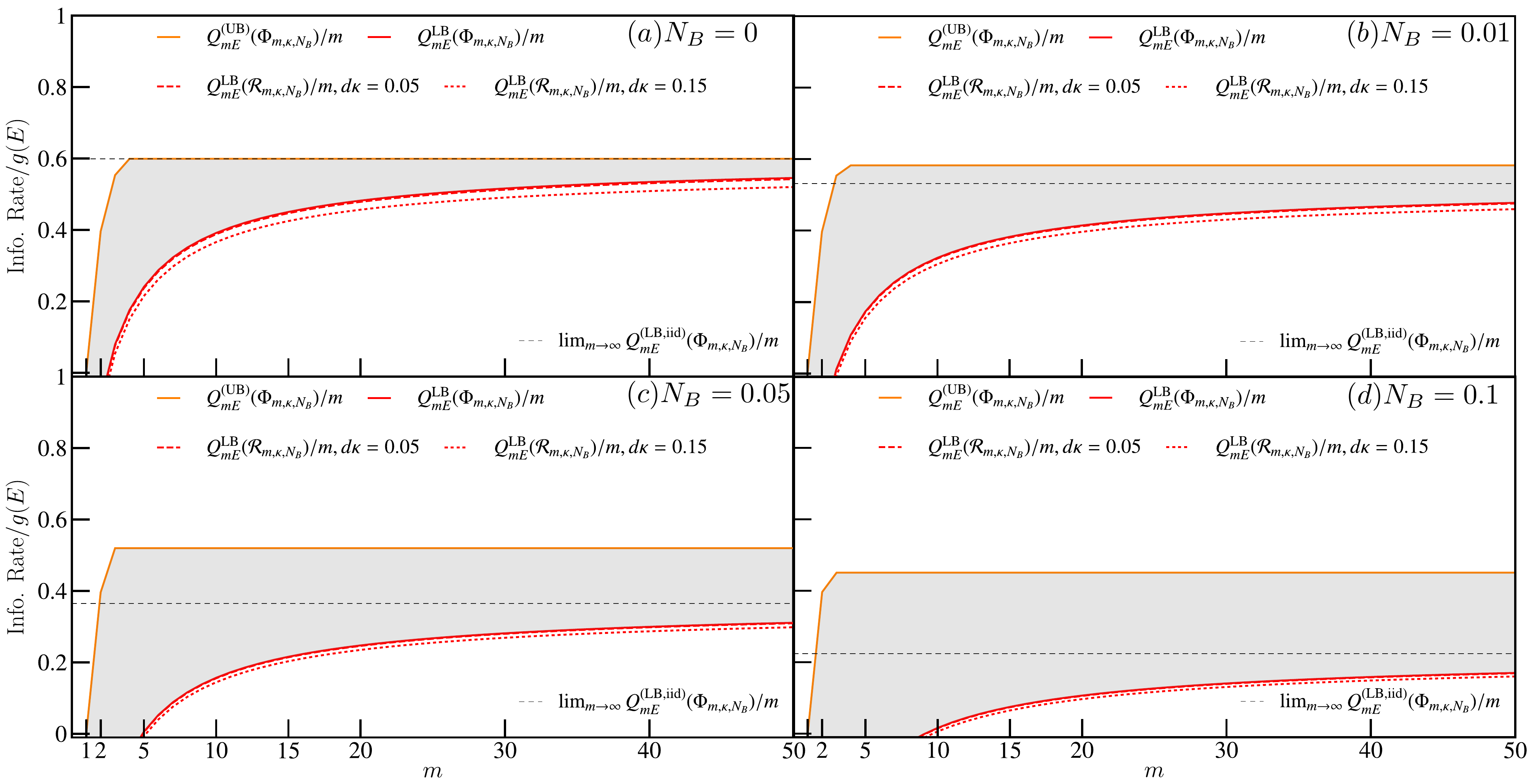}
\caption{Ratio of the information rate over $g(E)$, with an input energy per mode $E=1$. The quantum per mode $Q(\Phi_m)/m$ lies between the upper bound $Q_{mE}^{(\rm UB)}\left(\Phi_{m,\kappa,N_B}\right)/m$ (orange solid) and the lower bound $Q_{mE}^{\rm LB}\left(\Phi_{m,\kappa,N_B}\right)/m$ (red solid). The $m\to\infty$ limit of the lower bound is plotted in black dashed. The case with fading is plotted in red dashed ($d\kappa=0.05$) and red dotted ($d\kappa=0.15$). $\kappa=0.85,E=1$ and (a)$N_B=0$. (b)$N_B=0.01$. (c)$N_B=0.05$. (d)$N_B=0.1$.
\label{fig:CE_Q}
}
\end{figure*}

\begin{figure*}
 \centering
\includegraphics[width=0.5\textwidth]{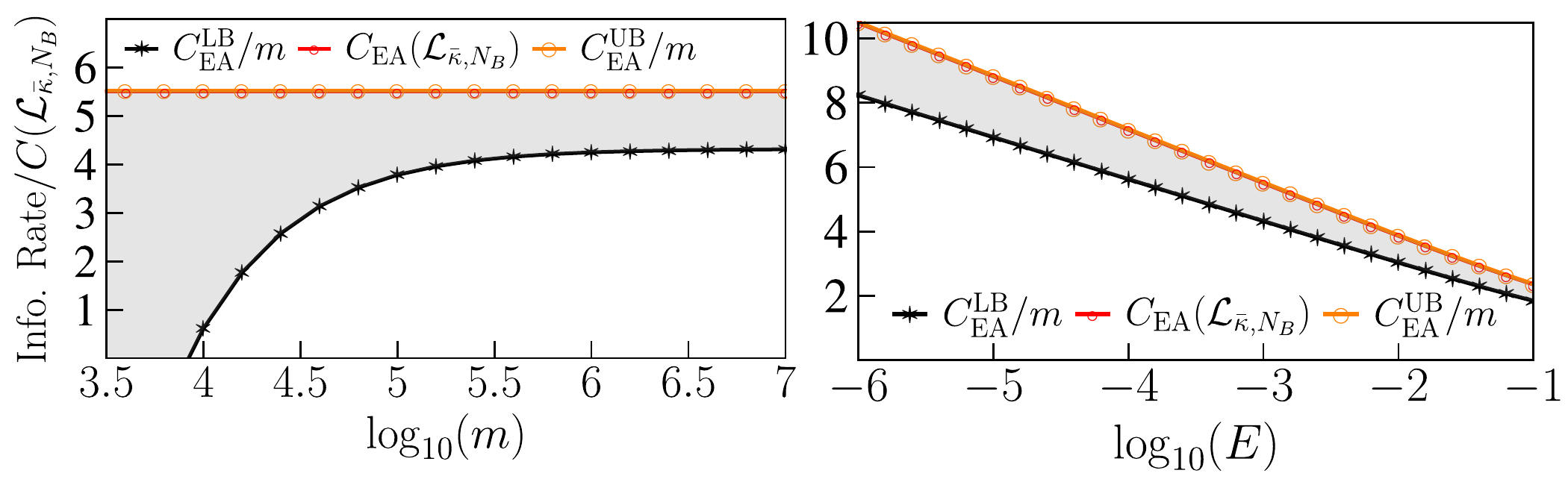}
\caption{
Ratio of information rate of a Rayleigh fading thermal-loss channel $\calR_{m,\bar{\kappa},N_B}$ over the classical capacity per mode upper bound $C(\mathcal{L}_{\bar{\kappa},N_B})$, with $\bar{\kappa}=0.1$. The EA capacity per mode $C_{\rm EA}(\calR_{m,\bar{\kappa},N_B})/m$ lies in the gray region, between the upper bound $C_{\rm EA}^{\rm UB}/m$ (orange circle) and the lower bound $C_{\rm EA}^{\rm LB}/m$ (black star). The estimated EA capacity $C_{\rm EA}(\calL_{\bar{\kappa},N_B})$ is also shown (red circle). (a) Energy fixed at $E=0.001$ per mode. (b) The $m\to\infty$ limit.  
\label{fig:CE_fading}
}
\end{figure*}

First, we begin with the EA classical capacity in comparison with the classical capacity.
We compare the exact lower bound $C_{\rm EA}^{\rm LB}/m$ in Eq.~\eqref{CEA_LB} and the asymptotic expression $C_{\rm EA}^{\rm LB, asym}/m$ in Eq.~\eqref{CEA_LB_asym}
in Fig.~\ref{fig:CE_noisy} for the noisy case, and find a good agreement. In the noisy case, we don't know the exact classical capacity of $\Phi_{m,\kappa,N_B}$, so we take the ratio of the information rate over $C(\mathcal{L}_{\kappa,N_B})\ge C(\Phi_{m,\kappa,N_B})/m$. 
Similar to the noiseless case, as we see in Fig.~\ref{fig:CE_noisy}, the lower bound converges to the upper bound $C_{\rm EA}(\calL_{\kappa,N_B})$ (see Eq.~\eqref{CE_UB}) quickly and revives the huge advantage over the HSW classical capacity when noise $N_B$ is large (e.g. sub-figure (a)). 
We also compare the phase encoding on TMSV in Fig.~\ref{fig:CE_noisy}. Indeed, we see in subfigure (a), when $N_B=10$ the accessible information lower bound per mode $\chi_{\rm LB}/m$ (red open circles) overlaps with the EA capacity lower bound (black stars); when $N_B$ is smaller, $\chi_{\rm LB}/m$ is lower than the EA capacity lower bound, however still provides an advantage over the HSW capacity. We also see a good agreement between the asymptotic accessible information lower bound $\chi_{\rm LB,asym}\left(\Sigma_{\bm \theta}\right)$ (red dashed) and $\chi_{\rm LB}/m$ (red open circles) when $m$ is not too small, as expected.

In Fig.~\ref{fig:CE_Q} we plot the results related to quantum capacity, as the ratio over $g(E)$, with an input energy per mode $E=1$. The quantum capacity per mode $Q(\Phi_m)/m$ lies between the upper bound $Q_{mE}^{(\rm UB)}\left(\Phi_{m,\kappa,N_B}\right)/m$ (orange solid, see Eq.~\eqref{QUB_final}) and the lower bound $Q_{mE}^{\rm LB}\left(\Phi_{m,\kappa,N_B}\right)/m$ (red solid, see Eq.~\eqref{Q_LB}). The $m\to\infty$ limit of the lower bound is plotted in black dashed. The case with fading (see Eq.~\eqref{Q_LB_fading}) is plotted in red dashed ($d\kappa=0.05$) and red dotted ($d\kappa=0.15$).

In presence of Rayleigh fading, we compare the EA capacity lower and upper bounds in Fig.~\ref{fig:CE_fading}, where we see a gap between them even at the $m\to\infty$ limit. The estimate of EA capacity $C_{\rm EA}(\calL_{\bar{\kappa},N_B})$ is very close to the upper bound. Most importantly, despite the fading, the logarithmically-diverging EA advantage persists. Note that here the $m\to \infty$ limit does not imply the long-term memory limit of compound channel~\cite{datta2007coding,datta2009classical,boche2017entanglement,boche2016entanglement,berta2017entanglement}.

\section{Additional details} 

\subsection{Evaluating the covariance matrix}
\label{cov_mat}
The covariance matrix of a TMSV with mean photon number $E$ is
\begin{align}
& 
{\mathbf{{\mathbf{\Lambda}}}}_{\rm TMSV} =
\left(
\begin{array}{cccc}
(2E+1) {\mathbf I}&2C_0{\mathbf Z}\\
2C_0 {\mathbf Z}&(2E+1){\mathbf I}
\end{array} 
\right),
&
\label{cov_TMSV}
\end{align}
where ${\mathbf I}$, ${\mathbf Z}$ are two-by-two Pauli matrices, and $C_0=\sqrt{E\left(E+1\right)}$ is the amplitude of the phase-sensitive cross correlation. 

The state $\calL_{x^2,N_B}\otimes \calI \left(\hat{\phi}_{\rm TMSV}\right)$ has the covariance matrix
\begin{align}
& 
{\mathbf{{\mathbf{\Lambda}}}}_{\rm TMSV}^{(x)} =
\left(
\begin{array}{cccc}
(2(x^2 E+N_B)+1) {\mathbf I}&2xC_0{\mathbf Z}\\
2xC_0 {\mathbf Z}&(2E+1){\mathbf I}
\end{array} 
\right).
&
\label{cov_TMSV_Lx}
\end{align}
Then the average state $\hat{\zeta}_{1,G}$ has the covariance matrix as
\begin{align}
& 
\braket{{\mathbf{{\mathbf{\Lambda}}}}_{\rm TMSV}^{(x)}}_{f(x)} =
\left(
\begin{array}{cccc}
(2(\bar{\kappa} E+N_B)+1) {\mathbf I}&2s_1C_0{\mathbf Z}\\
2s_1C_0 {\mathbf Z}&(2E+1){\mathbf I}
\end{array} 
\right).
&
\label{cov_TMSV_Lx_final}
\end{align}
Denote its symplectic eigenvalues as 
\be 
\mu_\pm=\left(\sqrt{(A+S)^2-4C^2}\pm (S-A)\right)/2,
\ee  
with the constants 
\ba 
A&\equiv& 2\left(\bar{\kappa}E+N_B\right)+1
\\
S&\equiv& 2E+1
\\
C&\equiv& 2s_1\sqrt{E(E+1)}.
\ea 
We can calculate its entropy as
\be 
S\left(\hat{\zeta}_{1,G}\right)=g(\frac{\mu_+-1}{2})+g(\frac{\mu_--1}{2}).
\ee 

\subsection{Proof of Lemma~\ref{lemma:CUB}}
\label{proof_lemma_supp}

\begin{proof}
We need to consider the regularization, 
\begin{align}
&\calC(\calR)=\lim_{n\to \infty}
\nonumber
\\
&\frac{1}{n}\max_{\{p_i,\hat{\rho}_i\}}\left[S\left(\calR^{\otimes n}\left(\sum_i p_i \hat{\rho}_i\right)\right)-\sum_i p_i S\left(\calR^{\otimes n}\left(\hat{\rho}_i\right)\right)\right]
\\
&\le S_{\rm max}-\lim_{n\to \infty} \frac{1}{n} \min_{\{p_i,\hat{\rho}_i\}} \sum_i p_i S\left(\calR^{\otimes n}\left(\hat{\rho}_i\right)\right)
\\
&\le S_{\rm max}-\lim_{n\to \infty} \frac{1}{n} \min_{\hat{\rho}} S\left(\calR^{\otimes n}\left(\hat{\rho}\right)\right),
\label{CIneq1}
\end{align}
where we have used $S\left(\calR^{\otimes n}\left(\sum_i p_i \hat{\rho}_i\right)\right)\le n\log_2(d)$ for $d$-dimensional case and $S\left(\calR^{\otimes n}\left(\sum_i p_i \hat{\rho}_i\right)\right)\le nmg(E^\prime)$ for the energy constrained infinite dimensional case, where the thermal state entropy $g(x)$ is concave in $x$.

Now we look at the second term
\begin{align}
&S\left(\calR^{\otimes n}\left(\hat{\rho}\right)\right)
\nonumber
\\
&=S\left(\left(\sum_k q_k \calR_k\right)^{\otimes n}\left(\hat{\rho}\right)\right)
\\
&=S\left(
\sum_{k_1,\cdots, k_n} q_{k_1}\cdots q_{k_n} \calR_{k_1}\otimes \cdots \otimes \calR_{k_n}
\left(\hat{\rho}\right)\right)
\\
&\ge 
\sum_{k_1,\cdots, k_n} q_{k_1}\cdots q_{k_n} S\left(\calR_{k_1}\otimes \cdots \otimes \calR_{k_n}
\left(\hat{\rho}\right)\right)
\\
&\ge 
\sum_{k_1,\cdots, k_n} q_{k_1}\cdots q_{k_n}
S_{\rm min}\left(\calR_{k_1}\otimes \cdots \otimes \calR_{k_n}\right)
\\
&= 
\sum_{k_1,\cdots, k_n} q_{k_1}\cdots q_{k_n}
\sum_{\ell=1}^n 
S_{\rm min}\left(\calR_{k_\ell}\right)
\\
&= 
n\sum_{k} q_k
S_{\rm min}\left(\calR_k\right).
\label{CIneq2}
\end{align}
Combining Ineq.~\eqref{CIneq1} and~\eqref{CIneq2}, we have the final result in Eq.~\eqref{C_UB_fading}.
\end{proof}

\end{document}